\documentclass{IEEEtran}

\usepackage[sort,compress,super]{cite}
\usepackage{amsmath,eqnarray}
\usepackage{graphicx}
\usepackage{amssymb}
\usepackage{tikz}
\usepackage{array}
\usepackage{url}
\usepackage{textcomp}
\usepackage{algorithmic}
\graphicspath{{pngimages/}}
\usepackage[caption=false]{subfig}
\usepackage{cite}
\usepackage{xcolor}

\newtheorem{theorem}{Theorem}
\newtheorem{corollary}{Corollary}[theorem]
\newtheorem{definition}{Definition}

\begin{document}

\markboth{E.Kulkarni, N.Sundararajan and S.Sundaram}{Design of Heli-quad for Full-Attitude Fault-Tolerant control}

\title{Heli-quad Design for Full-Attitude Fault-Tolerant control under Complete Failure of an Actuator }

\author{Eeshan Kulkarni$^{a,}$ , Narasimhan Sundararajan$^a$, Suresh Sundaram$^a$}


\maketitle

\begin{abstract}
This paper presents a reliable variable pitch propeller quadcopter with a cambered airfoil propeller called Heli-quad that achieves full-attitude control under a complete failure of one actuator. The idea of employing a cambered airfoil in the propeller blade plays a pivotal role in the full attitude control under the failure of an actuator. Experimental data shows that the cambered airfoil propellers generates significantly higher torque than symmetric airfoil propellers, enabling yaw control even under a complete failure of an actuator.  The theoretical analysis clearly indicates that Heli-Quad with three actuators is sufficient to provide full-attitude control. The proposed unified fault-tolerant controller consists of a  outer loop position tracking controller, a proportional-derivative inner loop attitude controller, and a novel neural-network-based reconfigurable control allocation scheme that computes the actuator commands. Experimentally validated propeller aerodynamic data has been used to train the neural network. High-fidelity software-in-the-loop simulations using the SIMSCAPE environment are carried out to analyze the Heli-quad’s performance. From the empirical result, the maximum tolerable delay in Fault Detection and Isolation (FDI) is 180msec. The results indicate that even under the complete failure of one actuator, the position tracking performance of the Heli-quad is closer to nominal conditions. 
\end{abstract}

\IEEEkeywords{Variable Pitch Propellers; Cambered airfoil; Fault-tolerant Control; Quadcopter full attitude control, Propeller aerodynamics}

\section{INTRODUCTION}
Recent technological developments in sensing and computation platforms have led to the design of new multi-rotor-based Vertical Take-Off and Landing (VTOL) Unmanned Aerial Vehicles (UAV), which are gaining more attention in a wide range of applications like firefighting  [\citen{hari20},\citen{hari18}], agriculture [\citen{agriculture}], search mission [\citen{ben_search}]. UAVs have recently found applications in geologic and construction mapping [\citen{geology,construction,senthil2021}]. Recently, quadcopters have been  used for aggressive maneuvering in a cluttered indoor environment  [\citen{landry}],  high-speed flight in GPS-denied indoor environments  [\citen{mohta}], and passenger-carrying vehicles [\citen{airtaxi}, \citen{pavel}]. In all the above applications, the robustness and reliability of the multi-rotor system are essential for a successful mission.

Actuators are prone to failure anytime during the mission due to unforeseen faults. In the vast majority of multi-rotors, the actuator consists of a fixed-pitch propeller driven by an electric Brush-Less DC (BLDC) motor. Unlike fixed-wing UAVs, in VTOL systems, actuators are the source of lift as well as propulsion. Therefore, robustness against actuator failure is particularly crucial. The following conditions can lead to a failure of a single actuator in a fixed-pitch multi-rotor UAV: a) partial loss of propeller blades [\citen{partiallossofblades}]; b) RPM (Revolutions Per Minute) of the propeller goes to zero (motor fails); c) complete loss of propeller blades; d) separation of the mounting arm along with the actuator from the main body. Here, the complete failure of an actuator is defined as a condition when it produces zero thrust and torque. Note that the last three conditions will result in the complete failure of an actuator.

 Control of multi-rotor UAVs under partial loss of control wherein the actual actuator output differs significantly from the desired actuator output is investigated in [\citen{partialfailure},\citen{Nguyenpartialfailure}]. Other commonly used multi-rotor systems such as Hexacopter and Octocopter can handle the complete failure of a single actuator with the help of their inherently redundant motors  [\citen{taes},\citen{hexacopter}]. However, hexacopter/octocopter results in a larger size, making them difficult to navigate in a cluttered environment and also they are expensive. One can also use a parachute to recover the UAV from a complete failure safely [\citen{parachute}]. However, in such cases, the mission must be aborted due to a lack of controllability. Under complete failure of one motor, reduced attitude control for quadcopters requires sacrificing of control over the yaw angle because of the non-existence of an equilibrium point [\citen{freddi}]. The reduced attitude control can also be achieved for the quadcopters suffering from two or three complete motor losses [\citen{mueller}]. In reduced attitude control conditions,  position references may be still tracked at the cost of very high yaw rate requirements. Hence, such a quadcopter’s use may not be possible for safety-critical payloads. Full attitude (with yaw angle) control for quadcopters (using fixed-pitch propellers) under a single actuator's complete failure is not dynamically possible. Alternative designs with minimal mechanical modification, such as the variable pitch propeller quadcopters, may be used for full-attitude control under the complete failure of one actuator.

Variable Pitch Propeller (VPP) application on the quadcopter have been in existence in the literature for about a decade. In VPP multi-rotors, the propeller pitch angle can be changed in addition to its RPM. The actual benefits of VPP over standard fixed-pitch actuators are high actuator bandwidth, generation of negative thrust, power optimisation, and sustained inverted flight were first presented in  [\citen{cutler}]. Apart from these, the fundamental capabilities of VPP quadcopters are still not fully explored. The redundancy in VPP quadcopters offers exciting research directions in the domain of actuator fault-tolerant control of quadcopters.

In addition to the actuator faults mentioned before for the fixed-pitch multi-rotors, actuators on VPP multi-rotors can also encounter pitch stuck fault wherein the pitch angle of the propeller blades is locked at/between certain angles and cannot be changed even after applying the control input. Fault-tolerant controller for pitch stuck fault on VPP quadcopters was developed in  [\citen{baldini}]. Analysis of such faults on centrally powered VPP quadcopters was presented in [\citen{wang}]. In these cases, the complete failure of a central power source will result in a total loss of controllability. Controllability analysis of full-attitude fault-tolerant VPP quadcopter was performed in [\citen{wang2020}], However, the propeller parameter values assumed by the authors are unreasonable for practical purposes and cannot be used in real applications. The aforementioned variable pitch quadcopters provide flexibility but do not provide full control authority under a complete failure of one actuator.  In the VPP quadcopters literature either a symmetric airfoil is used in the propeller blades [\citen{cutler},\citen{chipade}] or it is not mentioned. The symmetric airfoils cannot provide sufficient torque within the realistic motor RPM bounds to attain   full-attitude controllability.

In this paper, we present a new variable pitch propeller-based quadcopter with a cambered airfoil called Heli-quad (as its structure resembles a quadcopter and is actuated like a helicopter) that can provide the necessary controllability under the complete loss of a  single actuator. The Heli-quad has a similar design to that of a quadcopter, with four actuators placed equidistant from the centre along the diagonals of a square. However, the propellers in a  Heli-quad employs a cambered airfoil in its design and actuator controls both the collective pitch angle of the propeller and its rotation speed (RPM). The novel concept of using a cambered airfoil in the propeller blades is vital for providing full attitude controllability under a complete failure of one actuator. An intelligent controller is proposed to provide the necessary position tracking performance under the complete failure of a single actuator for the Heli-quad. Note, the Heli-quad does not provide full attitude controllability under multiple complete failures of actuators or under partial failure of an actuator(s). However, the partial failure cases can be handled by any of the related control methods mentioned in the above  literature.

The proposed control architecture includes a nonlinear quaternion-based outer-loop position controller and a proportional-derivative-based inner-loop attitude controller. Unlike fixed-pitch propellers, the relationship between thrust/torque generated in VPP and the pitch angle/RPM is highly nonlinear. Thus, to effectively map the nonlinear relationship between control inputs and the actuator commands, a neural network-based control allocation scheme is developed. The experimental data of the cambered propeller blade from the thrust bench has been used to train the neural network, which approximates the inverse functional relationship between the RPM and pitch of the propeller to thrust. The performance of the proposed controller and  the Heli-quad design are  evaluated in MATLAB  Simscape \textsuperscript{TM} environment \footnote{https://in.mathworks.com/products/simscape-multibody.html} . The custom-designed Heli-quad with a payload capacity of 50 grams is considered for the simulations. To simulate the complete actuator failure, one of the motors is switched off (RPM is commanded to go to zero) during the  hover conditions.
After failure, position references are changed to verify Heli-quad’s tracking capability with three fully working actuators. The same procedure is followed to simulate the actuator failure occurring while the Heli-quad is in a translation motion. In this case, one of the motors is switched off while the Heli-quad is moving forward \footnote{Given in supplementary material}, in other words, when it has a finite roll angle. A fault detection time of 50 milliseconds is assumed in both cases. It is observed that just after the failure, Heli-quad’s attitude increased to around 30deg (because of fault detection delay). However, the controller stabilizes it back to hover equilibrium in less than a second. In both cases, the fault-tolerant controller is compared with the nominal fault-free controller to validate  its  performance. The results clearly indicate that the proposed intelligent control can provide full-attitude control authority with the position tracking performance similar to the nominal fault-free conditions even under the complete loss of a single actuator. Simulations of other scenarios leading to the complete failure of an actuator, as mentioned before, won’t significantly affect the results. It was also found empirically, by conducting multiple simulations, that the maximum tolerable FDI fault detection time delay has enough margin (180 msec) over the assumed value of 50msec in the simulations.

 The main contributions of this article can be summarized as follows.
\begin{enumerate}
    \item Use of a cambered airfoil in the propeller blades to enable full-attitude control under the complete failure of an actuator. The difference between symmetric and cambered airfoil propeller is theoretically analysed and experimentally validated on a thrust measuring bench.
    \item A generalized control allocation method under an actuator failure that employs a neural network is developed and implemented in high-fidelity simulations.
    \item Maximum permissible FDI delay to ensure full-attitude stability after the complete failure of an actuator has been empirically estimated. 
\end{enumerate}

 The paper is organized as follows: The  Heli-quad design and experimental bench test results for the cambered and symmetric airfoils in the blade are described in section \ref{helidesign}. Controllability analysis and controller design for the Heli-quad is presented in section \ref{intelftcontroldesign}. High fidelity software-in-loop simulations are presented in section \ref{performanceeval}. Conclusions based on the study are given in section \ref{conclusion}.

\section{HELI-QUAD DESIGN} \label{helidesign}

In this section , we  describe  a new quadcopter design with variable pitch actuation and cambered airfoil to provide full control authority even under a complete failure of a single actuator. It is referred to as Heli-quad. The word Heli-quad is a blended word formed from a Helicopter and a Quadcopter. The Heli-quad's exterior appearance is similar to a conventional quadcopter, but the only difference is in the actuation. In a Heli-quad, one can manipulate the control thrust by varying each motor's speed like a quadcopter or change the propeller collective pitch angle like a helicopter's tail rotor or both. Heli-quad's actuator employs a brushless DC motor to control the propeller's speed and a servo motor to change the propeller blade pitch angle. Earlier works  on the use of variable pitch actuation in quadcopters make  use of symmetric air-foils in propeller blades [\citen{chipade2018advanced, cutler}]. These multi-copters can provide high maneuverability and inverted flight capabilities [\citen{cutler}], but full attitude controllability under the complete failure of a single actuator is not yet possible. The proposed Heli-quad overcomes the problem of control authority even under a single actuator failure. Further,  in the next section  an intelligent fault-tolerant control scheme to achieve full attitude control authority is presented.

Design requirements of a  Heli-quad consist of a  Micro Air Vehicle (MAV) class with an overall weight   lesser than 1 Kilogram, Payload carrying capacity of approximately 50 grams and full attitude controllability under the complete failure of a single actuator. In this section, first, the definition of complete failure is provided, and then the conditions to achieve equilibrium under the failure of a single actuator are discussed. Next, the propeller aerodynamics model is derived from  the first principles. Subsequently, the benefits of using a cambered air-foil over a symmetric air-foil in the propeller design are presented. This  section is concluded with the experimental bench test results.

\subsection{Definition of Complete Failure}
Control of multirotors under partial failure of motor is available in the literature [\citen{partialfailure},\citen{Nguyenpartialfailure}]. A parameter called loss of effectiveness is used to measure the degree of the failure. In such cases, actuators can generate thrust and torque but with less capacity. Complete failure is defined as the inability of an  actuator to produce any thrust or torque. It may happen as a result of a completely blown-off propeller or a fault in the BLDC motor. The complete failure complicates the stability and full attitude control authority of a  quadcopter.

\subsection{Static Equilibrium with Three Operating Actuators}
 \begin{figure}
 \begin{center}
 \centering
 \includegraphics [width=0.8\linewidth]{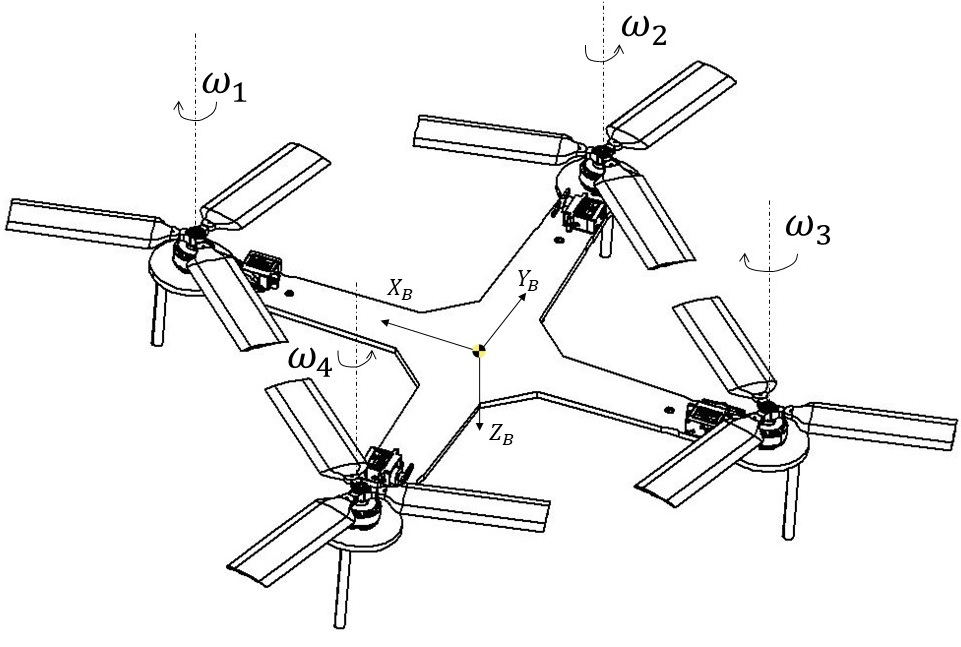}
\caption{Frame of references along with propeller rotation direction (subscript of $\omega$ is actuator number). The $4$th actuator is failed completely.}
\label{droneschematic}
\end{center}
\end{figure}
Static equilibrium or hover is possible in fixed-pitch quadcopter when the vehicle weight is balanced by the collective thrust produced by the four propellers. The  torque generated by two opposite working propellers is nullified by other sets of propellers. Complete failure of one actuator induces asymmetry in the design. In such conditions, a fixed-pitch quadcopter can't maintain hover equilibrium. This section shows that the proposed Heli-quad can qualitatively achieve hover equilibrium with only three working actuators.    Without any loss in generality, it is assumed that the actuator $4$ on the  Heli-quad shown in  Fig .\ref{droneschematic}  is not working   To maintain hover equilibrium, the following equations must be satisfied.
 \begin{equation}
     T_1 + T_3 = mg
     \label{t1+t3=mg}
 \end{equation}
 \begin{equation}
     T_1=T_3=\frac{mg}{2}
     \label{t1=t3=mg/2}
 \end{equation}
 \begin{equation}
    T_2=0
    \label{t2=0}
\end{equation}
\begin{equation}
\tau_2=\tau_1 + \tau_3
    \label{tau2=tau1+tau3}
\end{equation}
where, $T_i$ and $\tau_i$ are the thrust and torque produced by the $i^{th}$ propeller respectively, $mg$ is the weight of the Heli-quad. If total thrust-to-weight ratio of the vehicle is greater than two then it is possible to satisfy Eq.(\ref{t1+t3=mg}) and Eq.(\ref{t1=t3=mg/2})  The use of a variable pitch actuator can qualitatively satisfy conditions Eq.(\ref{t2=0}) and Eq.(\ref{tau2=tau1+tau3}) because of it has  the ability to produce zero thrust and non-zero torque simultaneously. However, propeller aerodynamics will play a pivotal role in the feasibility of satisfying the above conditions.

\subsection{Propeller Aerodynamics Model}
\begin{figure}
\begin{center}
\centering{\includegraphics [width=3.5in]{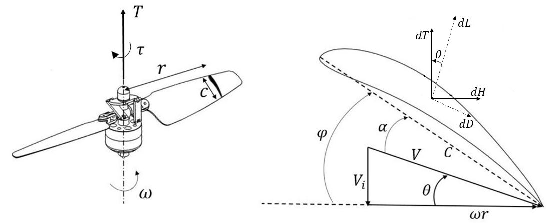}}
\caption{Airfoil section located at distance $r$ from axis of rotation (left), flow diagram for that airfoil section (right), $\phi$ is the pitch angle of propeller, $V_i$ is constant axial induced velocity, $V$ is the total velocity as seen by airfoil section, $\theta$ is flow angle and $\alpha$ is angle of attack of airfoil section.}
\label{bemt}
\end{center}
\end{figure}

The propeller aerodynamics is analyzed using the Blade Element Momentum Theory (BEMT) [\citen{shastry}].
In this paper minimal external wind is assumed, hence elemental thrust ($dT$) force acting on the airfoil section located at a distance $'r'$ from the center of rotation and rotating with angular velocity $\omega$ about it is given by Blade Element Theory as

 \begin{eqnarray}
  dT & = & dL cos(\theta) - dD sin(\theta)   \\
  & = &\frac{N_B}{2}\rho V^2 (C_l cos(\theta)- C_d sin(\theta)) c(r) dr
  \label{dtdl}
 \end{eqnarray}
where
 \begin{eqnarray}
     V^2 &=& V_i^2+ (\omega r)^2
     \label{vviwr} \\
  \alpha & = &\phi-\theta 
  = \phi-tan^{-1}\frac{V_i}{\omega r}  
  \label{alphaphi}
 \end{eqnarray}
 where, $dL$ and $dD$ are elemental lift and drag force acting on propeller blades (refer Fig. \ref{bemt}). These forces act perpendicular and parallel to the resultant air velocity vector respectively. The term $\rho$ is the air density (1.22 $Kgm^{-3}$),  $C_l=C_l(Re,\alpha)$ and $C_d=C_d(Re,\alpha)$ are the  lift and drag coefficients of airfoil used in the propeller, $Re$ is  the Reynolds number. Note, $c(r)$ is the chord length of section and $N_B$ is the number of blades. The elemental torque about the axis of rotation ($d\tau$) is given by
 \begin{equation}
  d\tau= rdH =  \frac{N_B}{2}\rho V^2 r (C_l sin(\theta)+ C_d cos(\theta)) c(r) dr
  \label{dtau}
\end{equation}\linebreak

In Eq.(\ref{dtdl}) there are two unknowns namely $dT$ and $V_i$. To solve these unknowns, momentum theory is used, which relates $dT$ in terms of $V_i$. The relationship is given by
\begin{equation}
dT=4\pi \rho F V_i^2 r dr\\
\label{dtmomentum}
\end{equation}
where \begin{equation}
 F=\frac{2}{\pi} cos^{-1}e^{-f} ,\qquad f={-\frac{N_B}{2} (\frac{1-r}{rsin\theta})}
 \label{tiploss}
\end{equation}
where, $F$ is prandtl's tip loss correction factor [\citen{leishman}]. Here, Eq.\eqref{dtdl} and Eq.\eqref{dtmomentum} are solved iteratively for each input $\phi$ and $\omega$  to find the corresponding unknown $V_i$. The known $V_i$ can be substituted back in Eq.(\ref{dtdl}) and Eq.(\ref{dtau}), and integrated over radius of the propeller to find the thrust and torque respectively.

Note that many combinations of control inputs $\phi$ and $\omega$ can generate the same desired thrust. High propeller pitch ($\phi$) can generate the desired thrust by keeping low motor RPM ($\omega$), and the same magnitude of thrust can be generated by setting $\phi$ to a lesser value but increasing $\omega$. In the interest of satisfying equations (\ref{t1+t3=mg})-(\ref{tau2=tau1+tau3}), optimal choice of control inputs $\phi$ and $\omega$ for actuator $1$ and $3$ would be the one that gives desired thrust with a minimum torque so that propeller $2$ can counteract it with less effort.
 
\subsection{Heli-quad Propeller Design} \label{helipropdesign}

In this section, the design of the propeller used in the Heli-quad is discussed. Airfoil used in the propeller plays a vital role in the yaw stability of a Heli-quad. To show it's importance  Eq.(\ref{dtdl}),(\ref{alphaphi}) and (\ref{dtau}) are simplified by assuming $\theta\approx 0$ or $V_i << \omega r$, Simplified BEMT equations can be written as
\begin{align}
  \alpha&\approx\phi  \label{alphaapproxphi} \\ 
  dT  \approx  dL  &\approx  \frac{1}{2}\rho  (wr)^2 C_l(re,\alpha) c(r) dr     \label{dtapproxdl}
\end{align}
Similarly,
 \begin{equation}
 d\tau\approx \frac{1}{2}\rho r  (wr)^2 C_d(re,\alpha) c(r) dr
 \label{dtauapprox}
 \end{equation}

Eq.(\ref{dtapproxdl}) and (\ref{dtauapprox}) encapsulate the essential information. It says that the thrust and torque generated by the propeller blades are heavily influenced by the lift and drag coefficients of the airfoil  respectively. The current literature on a multi-rotor with variable pitch propeller often use only symmetric air-foils in the propeller blades. Symmetric airfoils are efficient in producing lift in both directions but have a very low drag coefficient. This makes satisfying Eq.(\ref{tau2=tau1+tau3}) infeasible due to the upper limit on the rotor RPM. The very high RPM ($>$ 8000) requirement cause intense vibrations on the multi-rotor  [\citen{cutler}]. One can solve this problem by making use of cambered airfoils in the blade. Cambered airfoils have  very high drag coefficients at zero lift angles of attack which in turn can generate sufficient torque to satisfy Eq.(\ref{tau2=tau1+tau3}). Assuming a constant chord length, from Eq.(\ref{alphaapproxphi}),(\ref{dtapproxdl}),(\ref{dtauapprox}) it can be said that  
\begin{equation}
    \tau_{T=0} \propto Cd_{Cl=0}
    \label{propto}
\end{equation}
where $\tau_{T=0}$ is torque generated by propeller at zero thrust and $Cd_{Cl=0}$ is drag coefficient of airfoil when lift coefficient is zero.

\begin{figure}
\centering
\subfloat[Airfoils]{\includegraphics[width=\linewidth]{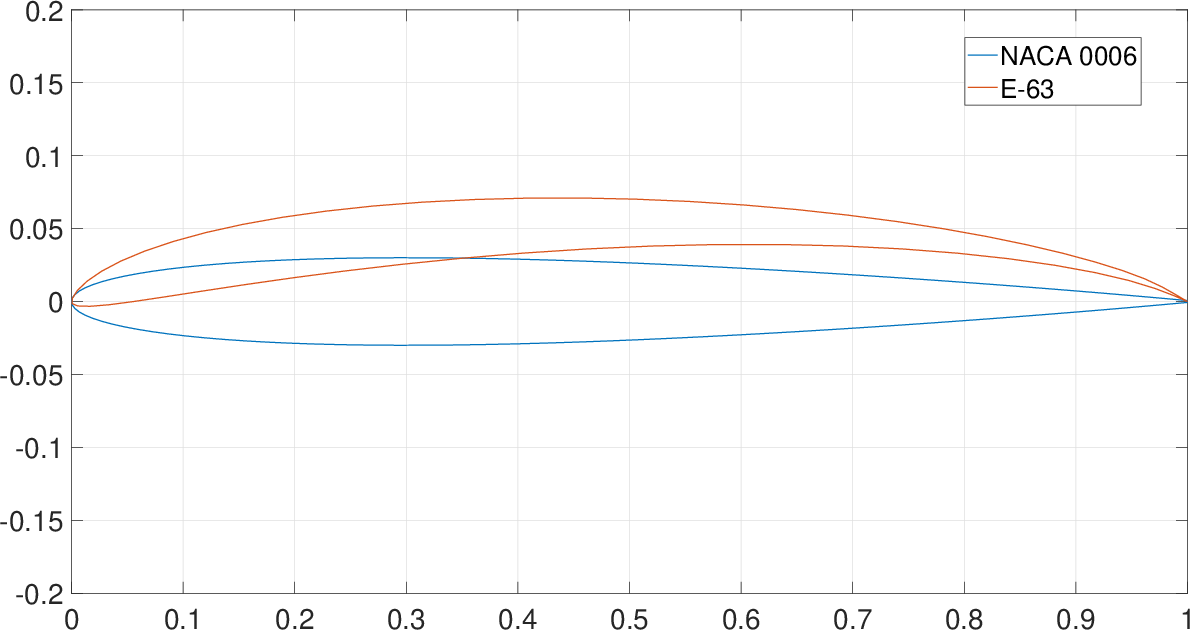}\label{airfoils}}
\\
\subfloat[Angle of attack vs coefficient of lift]{\includegraphics[width=\linewidth]{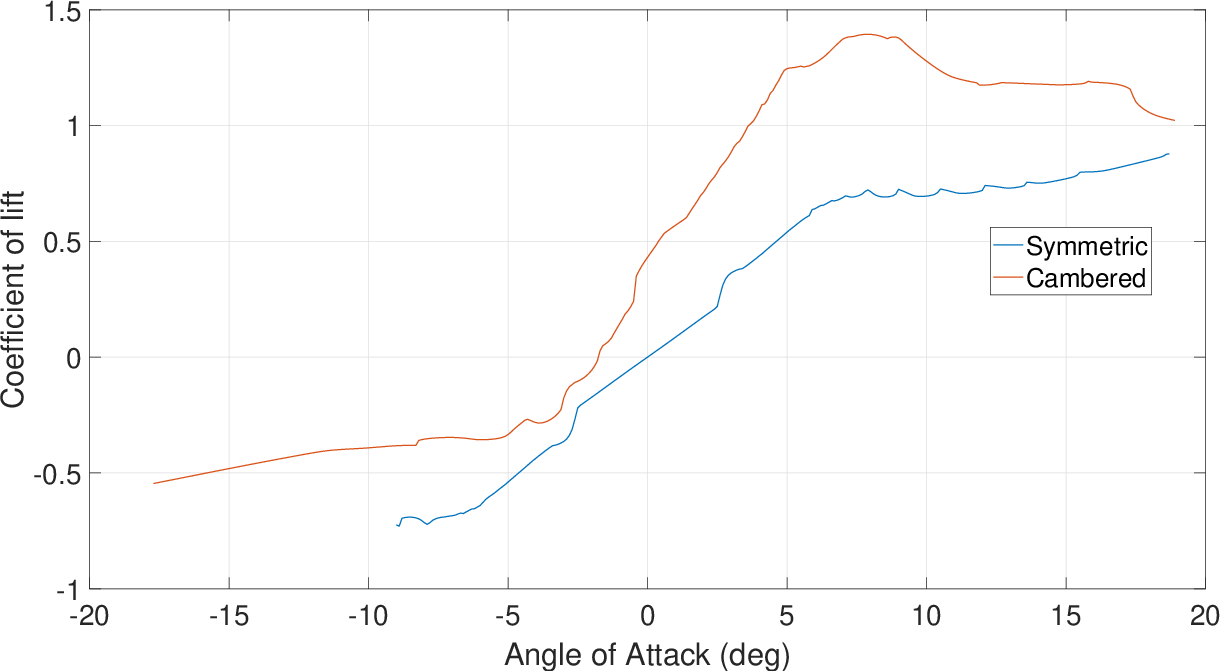}\label{alphavscl}}
\\
\subfloat[Coefficient of lift  vs Coefficient of drag]{\includegraphics[width=\linewidth]{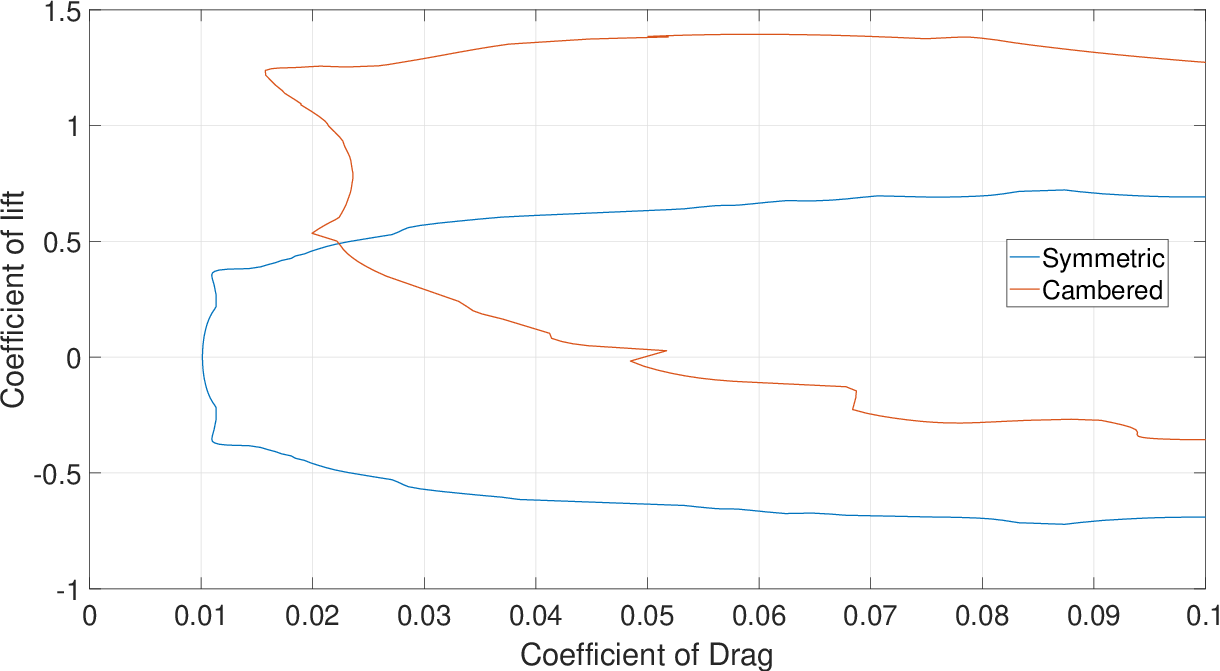}\label{cdvscl}}

\caption{(a) Airfoil shape comparison. (b) Lift characteristic comparison. (c) Lift vs Drag coefficient comparison. Airfoils are analysed at Reynolds number of 100,000 (Source : Xfoil code) }
\label{fig3}
\end{figure}

Fig. \ref{fig3} compares the geometry and aerodynamic characteristics of a typical symmetric airfoil (NACA 0006) used in the variable pitch propellers and a typical cambered airfoil (Eppler E-63) often used in commercially available fixed-pitch propellers\footnote{APC  propellers,2020.https://www.apcprop.com/technicalinformation/engineering}. Fig.(\ref{cdvscl}) shows that 
\begin{equation}
 Cd_{Cl=0,cambered} \approx 4.5 \times Cd_{Cl=0,symmetric}   
\end{equation}
therefore from Eq.(\ref{propto}), for same $\omega$
\begin{equation}
    \tau_{T=0,cambered} \approx 4.5 \times \tau_{T=0,symmetric}
\end{equation}

\begin{figure}  
\begin{center}
\centering
\includegraphics[width=0.7\linewidth]{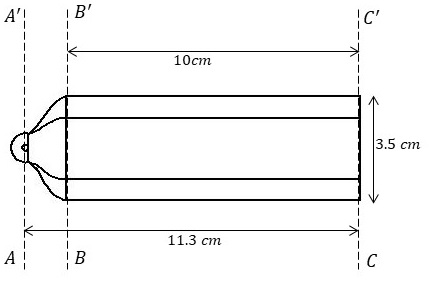}
\caption{ Geometry of the propeller blade.}
\label{bladegeometry} \end{center}
\end{figure}

\begin{figure*}
 \begin{center}
 \centering
\includegraphics [width=0.95\linewidth]{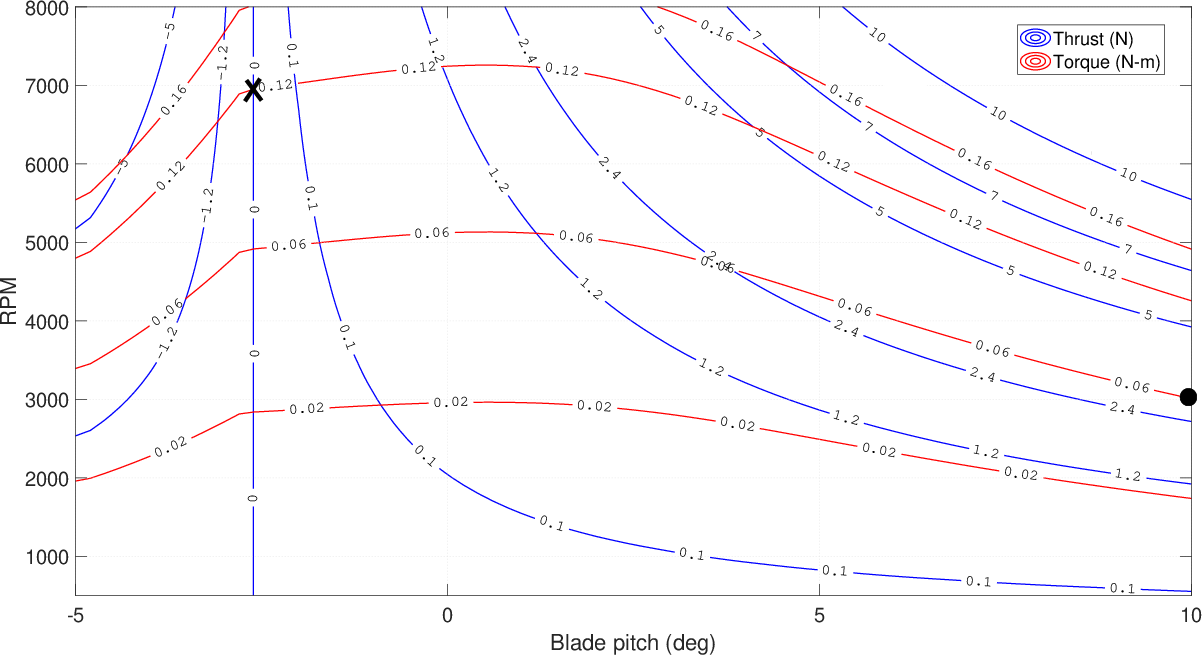}
\caption{Contour lines of thrust and torque produced by single actuator, propeller pitch angle and motor RPM can move anywhere in this plane. Cross represents equilibrium control input for actuator 2 when 4$^{th}$ one fails, solid circle represents the control input for actuator 1 and 3. Assumed mass of Heli-quad is 600 grams. Roll control is possible because of variable pitch actuation which can generate negative thrust.}
\label{thrusttorque}
 \end{center}
\end{figure*}

Cambered airfoil propeller gives $4.5$ times more torque at zero thrust than the symmetric airfoil propeller, while at higher thrusts there is not much difference in torque produced because of smaller difference in $C_d$. This characteristic of cambered airfoil helps in satisfying Eq.(\ref{t2=0}) and Eq.(\ref{tau2=tau1+tau3}).
Fig (\ref{bladegeometry}) shows the topography of the propeller blade used in Heli-quad. Note that E-63 low Reynolds number airfoil is used in the blades. Three such blades are used on a single rotor (Tri-blade) to increase the yaw controllability by increasing maximum $\tau_{T=0}$. BEMT data for this rotor is plotted in Fig (\ref{thrusttorque}). Mass of the Heli-quad plays a critical role in yaw attitude controllability, as mass goes down the 'cross' in the Fig (\ref{thrusttorque}) slides vertically downwards giving more yaw control authority. It should be noted that using cambered airfoil in the blades will deteriorate the negative thrust generation capability because of the asymmetry in Fig (\ref{alphavscl})  making it difficult to sustain inverted flight, however it is not of interest for this paper.
 
\begin{figure*}
\begin{center}
\centering
\includegraphics [width=0.6\linewidth]{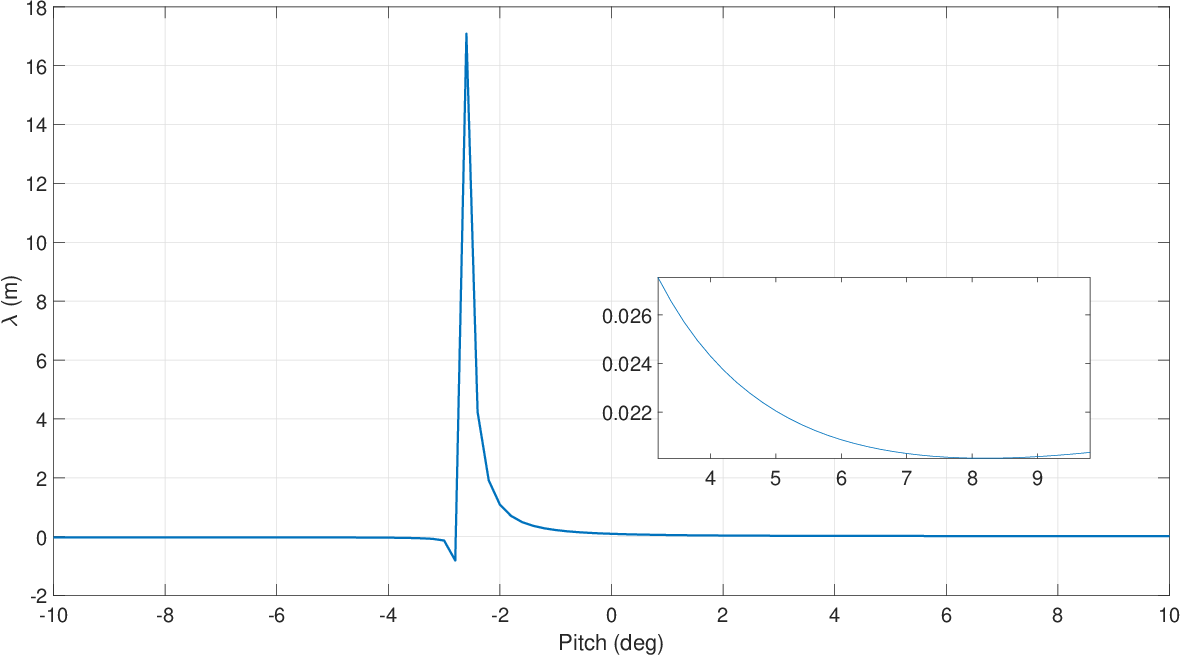}
\caption{Variation of torque to thrust ratio of single actuator with pitch angle. Insight shows it doesn't vary much at higher pitch angles }
\label{pitchlambda}
\end{center}
\end{figure*}

 A parameter $\lambda$ which is the ratio of torque to thrust produced by propeller is denoted as
\begin{equation}
    \lambda=f_{\lambda}(\phi)= \frac{\tau(\phi,\omega)}{T(\phi,\omega)}
\end{equation}
Note, the parameter $\lambda$ is invariant of RPM,it just depends on $\phi$. Function $f_{\lambda}$ for Heli-quad's propeller is plotted in Fig (\ref{pitchlambda}). In the figure it can be seen that $\lambda$ takes a very high value near $\phi_{T=0}$ because of very high torque and minimal thrust. $\phi_{T=0}$ is the pitch angle at which thrust produced is zero, it comes out to be around -2.6 degree for Heli-quad's propeller.

 \subsection{Experimental Bench Test Results} \label{experimental}
 
\begin{figure}
\centering
\subfloat[]{\includegraphics [width=0.5\linewidth]{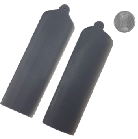}\label{planform}}
\\
\subfloat[]{\includegraphics [width=\linewidth]{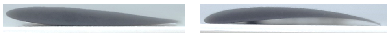}\label{crosssection}}
\caption{(a) Planform of both the blades is same (b) Symmetric (Left) and cambered( Right) airfoils used in the blades }
\label{propphoto}
\end{figure}

 In this subsection, cambered airfoil and symmetric airfoil propellers are tested on a thrust stand to evaluate the existence of a hover equilibrium. Experimental validation of BEMT is also presented. The airfoil used in the propeller plays a crucial role in the yaw control under a complete failure of an actuator. Two sets of propeller blades, one with a symmetric airfoil (NACA 0006) and the other with a cambered airfoil (E-63) as mentioned in section \ref{helipropdesign} were manufactured and tested for the generated thrust and torque values at different pitch angles and angular velocities (RPM's).  Fig.\ref{propphoto} shows the planform and the cross-section of two different propeller blades that were tested. The propeller blade dimensions are the same as shown in Fig.\ref{bladegeometry}. 
 
\begin{figure}
\centering
\subfloat[]{\includegraphics [height=1.8in,width=0.75\linewidth]{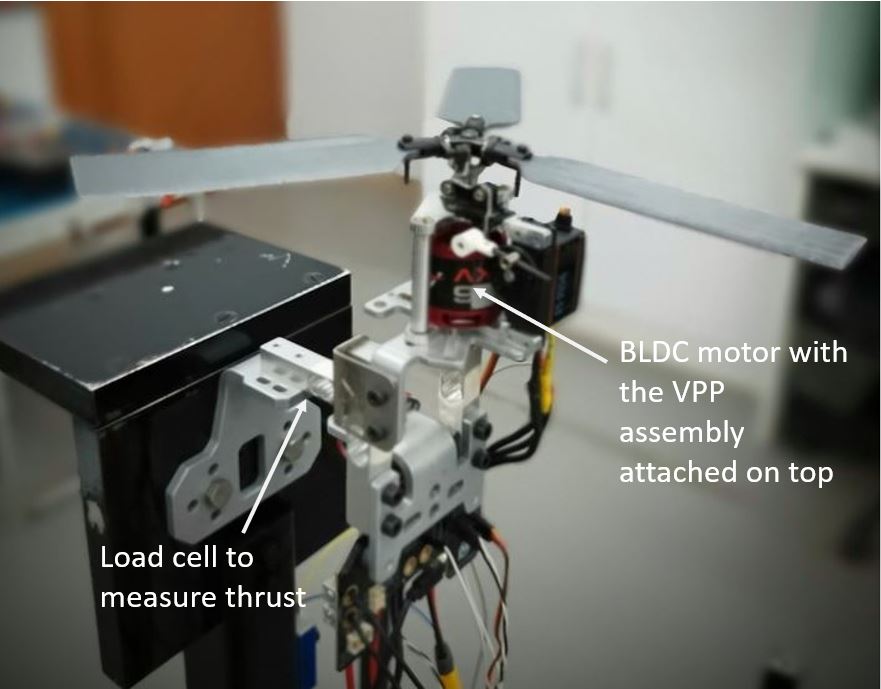}\label{test}}
\\
\subfloat[]{\includegraphics [height=1.8in]{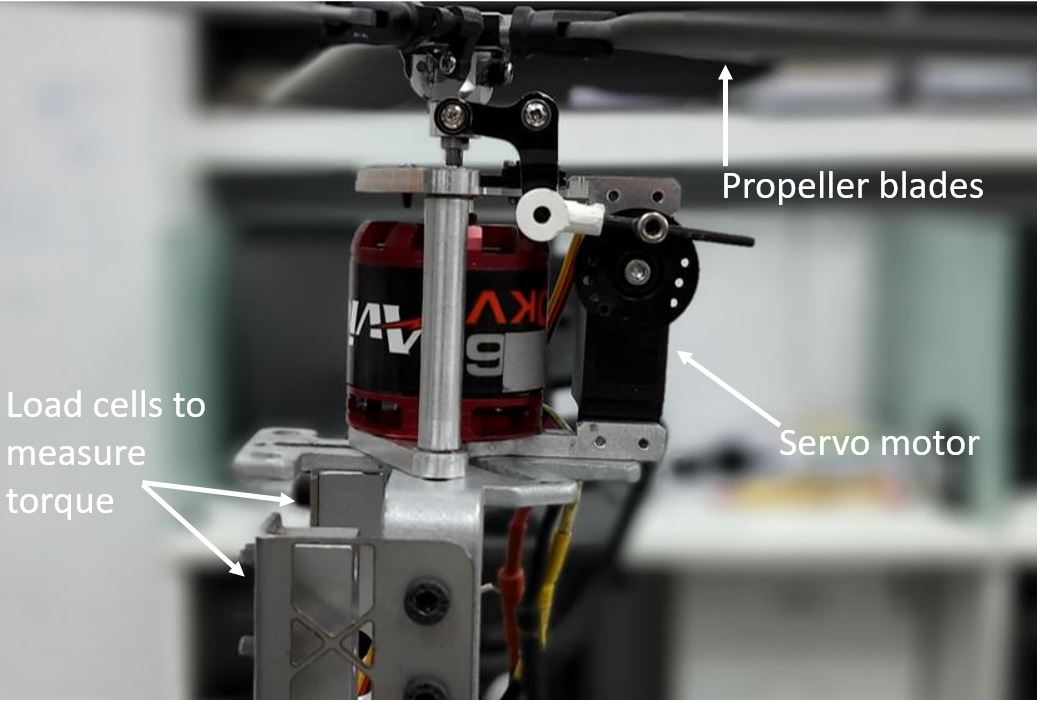}\label{mechanism}}
\caption{(a) Heli-quad's actuator mounted on a load cell (b) Side view of the variable pitch mechanism }
\label{teststand}
\end{figure}

 Stereolithography (SLA) additive manufacturing technique was used to manufacture the propeller blades. This 3D printing method was chosen because of it’s ability to print high tolerance, high strength, and smooth surface finish geometries. The blades were printed in a ’Formlabs Form 3+’ 3D printer and using their ’Tough 2000’ resin material \footnote{https://formlabs.com/store/tough-2000-resin/}. Fig.\ref{teststand} shows the Heli-quad’s actuator mounted on the load cell. An outrunning BrushLess-DC motor was modified by replacing it’s original shaft to a longer one for mounting the collective pitch mechanism. This mechanism was obtained from the tail rotor of a commercially available radio control helicopter. A high torque digital servo motor was used to vary  the pitch of the blades. RPM of the BLDC motor was controlled by sending appropriate PWM commands to the Electronic Speed Controller (ESC). $12$ volt regulated DC supply was fed as the input to the ESC. Specific details of the hardware used in the experiments are given in Table \ref{hardware}. 
\begin{table}
\centering
\caption{Hardware specifications}\label{hardware}
{\begin{tabular}{@{}lcccr@{}}
\hline
 Part & Specifications \\
\hline
 BLDC Motor & Avionic PRO KV1050 \\
Servo Motor & Emax ES9257\\
VPP mechanism & 450 size 3 blade\\
Load cell & RcBenchmark 1585\\
ESC & Emax BLHeli 30A \\
\hline
\end{tabular}}
\end{table}
 
 To validate the existence of hover equilibrium for the Heli-quad, the thrust and torque values generated by the propeller blades were recorded for various RPMs at 10-degree pitch angle and zero-thrust pitch angle, as discussed in section \ref{helipropdesign}. To estimate the zero-thrust pitch angle, the PWM signal given to the motor was fixed at a constant value, and the propeller pitch varied until the thrust generated was near zero. The values were obtained by averaging the data over 100 samples after the system reached a steady state.

 \begin{figure}
\centering
\subfloat[RPM vs Thrust]{\includegraphics[width=\linewidth]{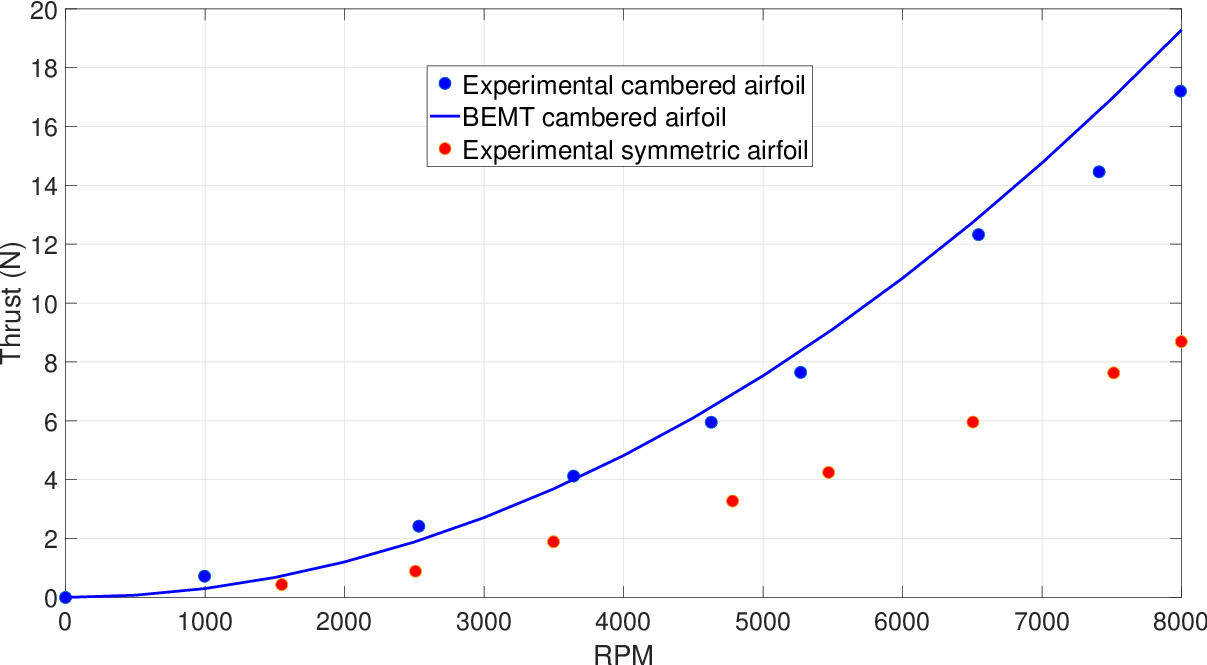}\label{10degthrust}}
\hfill
\subfloat[RPM vs Torque]{\includegraphics[width=\linewidth]{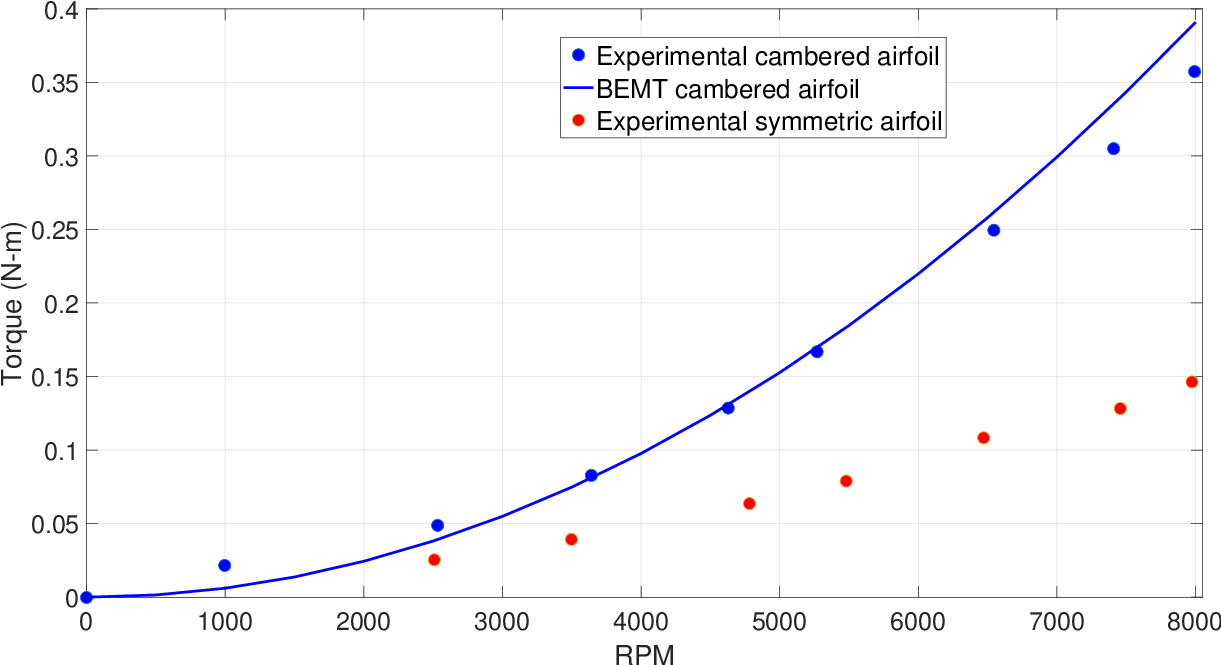}\label{10degtorque}}
\\[3.5ex]
\subfloat[Thrust vs Torque ]{\includegraphics[width=\linewidth]{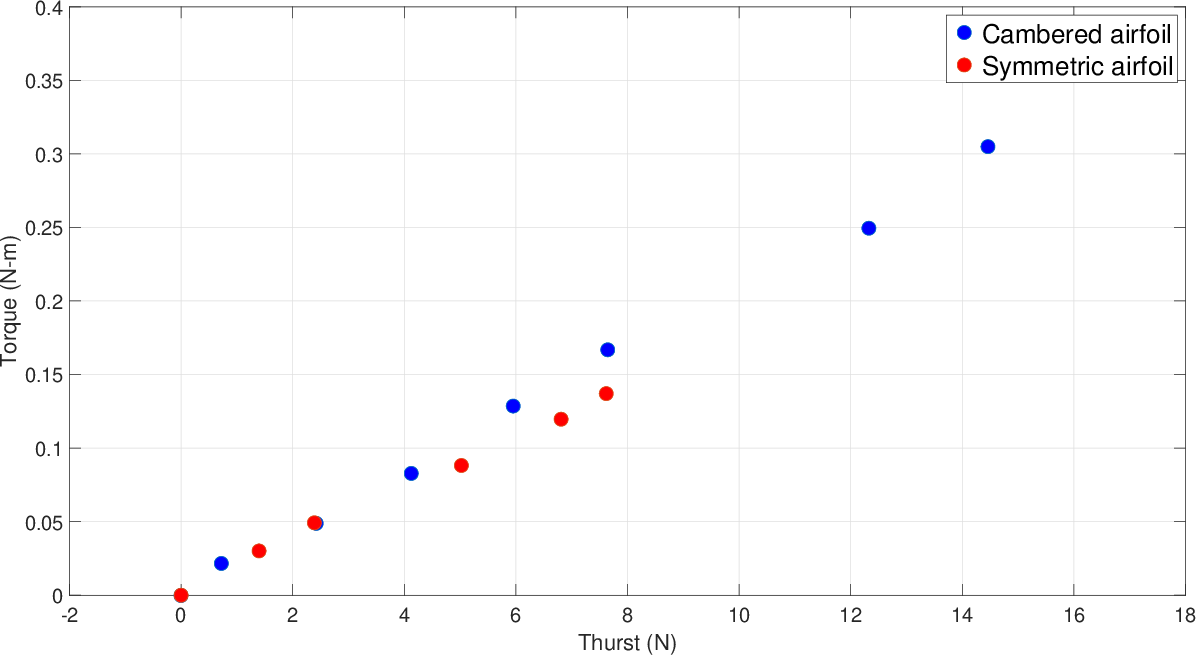}\label{tvsq}}
\caption{Propeller thrust and torque data for 10 degree pitch angle}
\label{10deg}
\end{figure}

Fig.\ref{10deg} shows the thrust and torque data as the function of rotor RPM for both the symmetric and cambered airfoil propellers at $10$ degrees pitch angle. The BEMT data for the cambered airfoil propeller is also overlaid for the theoretical validation. As predicted by BEMT, it can be seen that the cambered airfoil propeller generates significantly more thrust and torque than the symmetric airfoil propeller. However, as seen from  Fig.\ref{tvsq}  the difference between the torque produced at a given thrust doesn’t vary much between two airfoils. Hence, to maintain the hover equilibrium, the torque that needs to be generated by the actuator opposite to the failed motor will not differ significantly between these two propellers.

\begin{figure}
\centering
\subfloat[RPM vs Thrust]{\includegraphics[width=\linewidth]{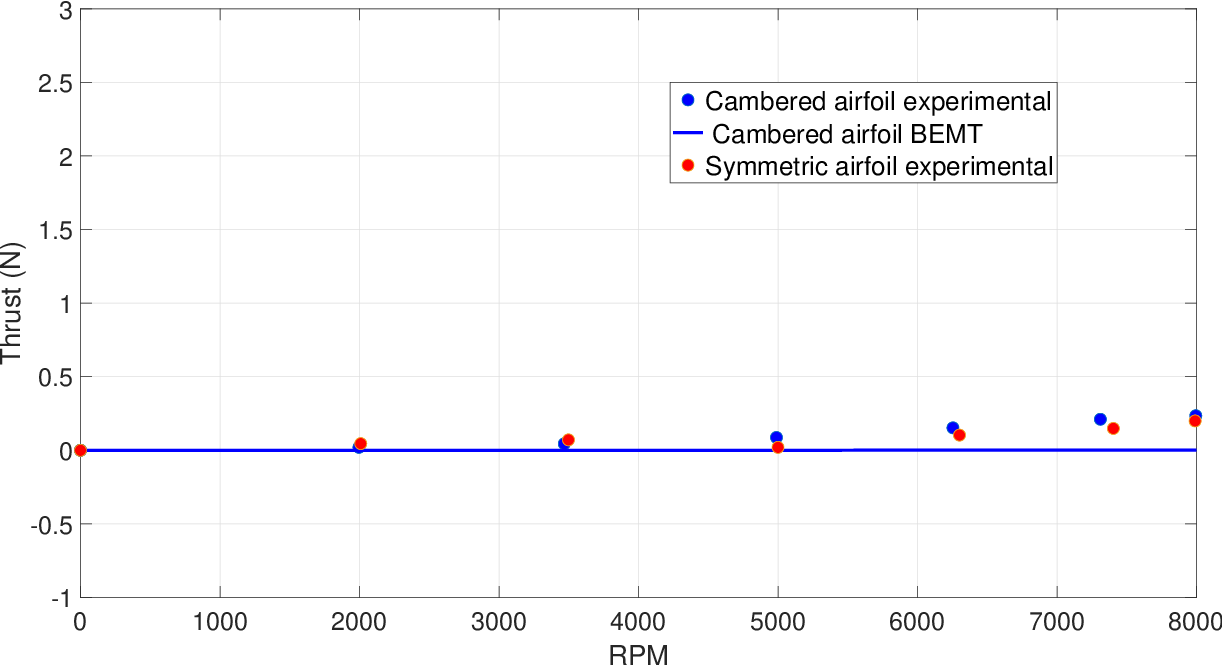}\label{0degthrust}}
\\
\subfloat[RPM vs Torque ]{\includegraphics[width=\linewidth]{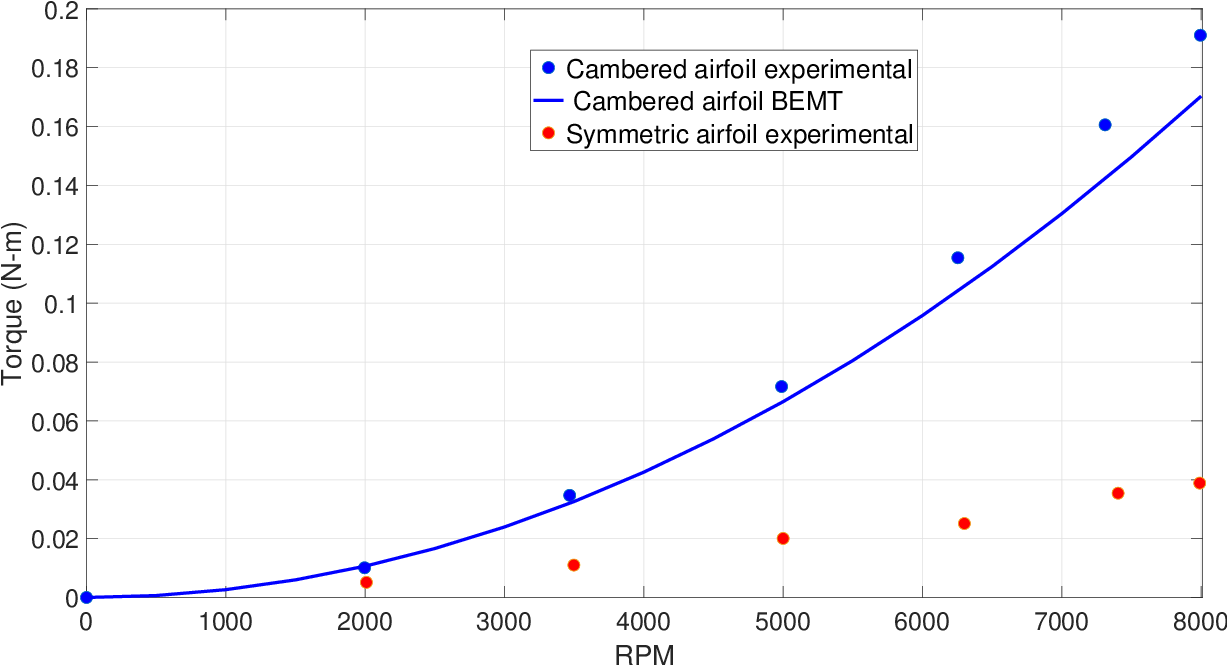}\label{0degtorque}}
\caption{Propeller thrust and torque data for zero thrust pitch angle ($\phi_{T=0}$)}
\label{0deg}
\end{figure}

Fig.\ref{0deg} shows the thrust and torque generated by two propellers at zero thrust pitch angle. The measured zero-thrust pitch angle is close to $0$ degrees for symmetric airfoil propellers and close to -3 degrees for cambered airfoil propellers. It can be seen from Fig.\ref{0degthrust} that the thrust generated at these pitch angles is insignificant, almost zero. Fig.\ref{0degtorque} validates a very important result. As predicted by BEMT, the cambered airfoil propeller generates a  torque almost 4.5 times that of symmetric airfoil propellers at a given RPM. This enables the hover equilibrium for the Heli-quad with realistic weights. As mentioned in [\citen{cutler}, it was observed that beyond 8000 RPM ( $>$ 85\% throttle), the vibrations are too high to operate safely. Hence, from Fig.\ref{tvsq} and Fig.\ref{0degtorque}, it can be estimated that the maximum weight for the Heli-quad that can achieve hover equilibrium is around 815 grams when using a cambered airfoil propeller. In contrast, when the symmetric airfoil propeller is used, the maximum weight drops to around $200$ grams, which is impractical.

The theoretical BEMT data is compared  with the experimental data for both the propellers as shown with the solid line in Fig.\ref{10deg} and Fig.\ref{0deg}. The indicated error is within 10\% for all curves (except fig.\ref{0degthrust}, where the measure is not valid).  Various sources like uncertainties in the aerodynamic parameters, inaccuracies in measuring the pitch angle and manufacturing defects contribute to this error. BEMT data augmented with such uncertainties has been used for the Heli-quad high fidelity simulations discussed below.

\section{INTELLIGENT FAULT-TOLERANT CONTROLLER DESIGN FOR A HELI-QUAD} \label{intelftcontroldesign}
 
In this section, an intelligent fault-tolerant control scheme for the Heli-quad that can handle a complete failure of a single actuator is presented. The control scheme employs a typical outer-loop position tracking controller and an inner-loop attitude control scheme similar to that in  [\citen{cutler}]. A Neural-Network based intelligent reconfigurable control allocation scheme is presented to handle both nominal and failure conditions. First, the controllability analysis under a faulty actuator is presented, then the outer and inner loop controllers are described followed up with, the reconfigurable control allocation scheme is presented.

\subsection{Controllability analysis} \label{controlanalysis}

In this section, the controllability analysis under a faulty actuator is presented. Null-controllabilty analysis of multirotors which relates to it's ability to reach static hover is also discussed.  

\begin{definition} [\citen{kim}]
 Consider a nonlinear system $\dot{x}=h(x, u)$, where $x(t) \in \mathbb{R}^{n}$ is a state vector, $u(t) \in \Omega \subset \mathbb{R}^{m}$ is a control input bounded by a restraint set $\Omega$, and $h$ is in $C^{1}$. The system is locally $\Omega$-null controllable if there exists an open set $\mathcal{V} \subset \mathbb{R}^{n}$ containing the origin such that any $x_{0} \in \mathcal{V}$ at time $t_0$ can be controlled to $x_{f}=0$ in finite time $t_f<\infty$ by some controller satisfying $u(t) \in \Omega$ for all time $t \in\left[t_{0}, t_{f}\right] .$ The system is globally $\Omega$-null controllable if $\mathcal{V}=\mathbb{R}^{n}$.
\end{definition}

Theorem \ref{theorem} and Corollary \ref{corollary} from [\citen{kim}] provide conditions for a system to be globally null controllable. Readers are referred to [\citen{nullcntrl}] for their details and proofs.

\begin{theorem} [\citen{kim}]
Consider a system $\dot{x}=h(x, u)$. Suppose there exist a scalar function $V(x): \mathbb{R}^{n} \rightarrow \mathbb{R}$, and a vector function $U(x): \mathbb{R}^{n} \rightarrow \mathbb{R}^{m}$ in $C^{1}$ such that \\
(a) $V(x)=0$ if and only if $x=0$, and $V(x)>0$ otherwise; \\
(b) $\lim _{\mid x \| \rightarrow \infty} V(x)=+\infty$; \\
(c) $U(\boldsymbol{x}) \subset \Omega$; \\
(d) $\dot{V}<0$ for $x \neq 0$, and $\dot{V}=0$ for $x=0$. \\
Then the system is globally asymptotically stable about the origin with the controller $u(t)=U(x(t)) \subset \Omega$ on $0 \leq t<\infty .$
If, in addition to $(d), \dot{V} \leq-\beta V$ holds for some $\beta>0$, then the system is exponentially stable.
\label{theorem}
\end{theorem}

\begin{corollary} [\citen{kim}]
 Consider the system in Theorem 1 and assuming that there exists $V(x)$ and $U(x)$ satisfying necessary conditions. If the followings also hold,\\
(e) $h(0,0)=0$;\\
(f) $u=0$ is in the interior of admissible control input set $\Omega$; where $u=0$ is the equilibrium control input.\\
(g) $\operatorname{rank}\left[\mathbf{B}, \mathbf{A B}, \cdots, \mathbf{A}^{n-1}\\ \mathbf{B}\right]=n$, where $\mathbf{A}=\frac{\partial h(0,0)}{\partial x}$ and $\mathbf{B}=\frac{\partial h(0,0)}{\partial u}$ \\, then the domain of null controllability for the system is $\mathbb{R}^{n}$.
\label{corollary}
\end{corollary}

Fault in the actuator will shrink the admissible control input set $\Omega$, making it difficult to satisfy condition $(f)$ in corollary \ref{corollary}. Condition $(f)$ can be satisfied only by employing cambered airfoil in the propeller design as shown in Fig (\ref{thrusttorque}). However, for the Heli-quad to be globally null controllable condition $(g)$ should also hold true, this is validated in the next section.

\subsection{Linear controllability analysis}
The dynamics of Heli-quad like any other quadcopter is given by,

\begin{multline}
    \dot{\mathbf{x}}=\left[\begin{array}{c}
\dot{x} \quad \dot{y} \quad \dot{z} \quad \dot{{\phi_h}} \quad \dot{\theta_h} \quad \dot{\psi_h} \quad \dot{p}  \quad \dot{q} \quad \dot{r} \quad \dot{u} \quad \dot{v} \quad \dot{w}
\end{array}\right]^T  =h(\mathbf{x},\mathbf{u}) \\
= \bigg[ u \quad v \quad w \quad p+qs_{\phi_h}t_{\theta_h}+rc_{\phi_h}t_{\theta_h} \quad qc_{\phi_h} - rs_{\phi_h}  \\ \quad qs_{\phi_h}sec\theta_h + rc_{\phi_h}sec\theta_h \quad qr \frac{J_{zz}-J_{yy}}{J_xx} + u_2 \frac{l}{J_{xx}}  \quad \\ \qquad  rp \frac{J_{xx}-J_{zz}}{J_yy} + u_3 \frac{l}{I_{YY}} \quad pq \frac{J_{yy}-J_{xx}}{J_zz} + u_4 \frac{l}{J_{zz}} \\\quad \frac{(c_{\psi_h} s_{\theta_h} c_{\phi_h}+s_{\psi_h} s_{\phi_h}) u_{1}}{m}  \quad \frac{(s_{\psi_h} s_{\theta_h} c_{\phi_h}-c_{\psi_h} s_{\phi_h}) u_{1}}{m} \\\quad \frac{c_{\theta_h} c_{\phi_h)} u_{1}}{m}   - g
\bigg]^T
\end{multline}
where, $[x \ y\  z]^T$ is the position vector of centroid of the Heli-quad in inertial frame, $[\phi_h \ \theta_h \ \psi_h]^T$ are the euler angles, $[p \ q \ r]^T $ is the body frame angular velocity vector, $[u \ v \ w]^T$ are linear velocities in body frame. $s_x=sin(x)$, $c_x=cos(x)$. It is assumed that actuator $4$ fails, then control input vector can be written as
\begin{equation}
\mathbf{u} =
\begin{bmatrix}
u_1\\ u_2\\ u_3\\ u_4
\end{bmatrix}=
\begin{bmatrix}
T_{\Sigma}\\
\tau_x\\
\tau_y\\
\tau_z\\

\end{bmatrix}=
\begin{bmatrix}
T_1+T_2+T_3\\
-T_2 l\\
(T_1-T_3)l\\
-\tau_1+\tau_2-\tau_3\\

\end{bmatrix}
\end{equation}

Note, because of variable pitch, $T_2$ can take both positive and negative values, therefore roll control authority can be achieved even if one rotor fails. As thrust and torque are highly nonlinear functions of $\phi$ and $\omega$, closed form solution of equilibrium actuator commands is not possible therefore control input vector is left in terms of $T$'s and $\tau$'s.
At hover equilibrium, state vector $\mathbf{x}$ and control input $\mathbf{u}$ can be written as,
\begin{align}
\mathbf{x}_{eq} &=
\begin{bmatrix}
x_0 \quad y_0 \quad z_0 \quad 0  \quad 0 \quad  \psi_0 \quad 0 \quad 0 \quad 0 \quad  0 \quad 0 \quad 0
\end{bmatrix}^T \nonumber \\
\mathbf{u}_{eq} &=
\begin{bmatrix}
mg \quad 0 \quad 0 \quad 0
\end{bmatrix}^T
\end{align}

Linearizing dynamics about the hover equilibrium point,
\begin{equation}
    A= \frac{\partial h(\mathbf{x},\mathbf{u})}{\partial \mathbf{x}}|_{(\mathbf{x_{eq}},\mathbf{u_{eq}})} ;\
    B=\frac{\partial h(\mathbf{x},\mathbf{u})}{\partial \mathbf{u}}|_{(\mathbf{x_{eq}},\mathbf{u_{eq}})} 
\end{equation}
 $A \in \mathbb{R}^{12\times12}$ and $B \in \mathbb{R}^{12\times4}$ are given by \\
\begin{equation}
A=\left[\begin{array}{llll}
0_{3 \times 3} & 0_{3 \times 3} & 0_{3 \times 3} & \mathbf{I}_{3 \times 3} \\
0_{3 \times 3} & 0_{3 \times 3} & \mathbf{I}_{3 \times 3} & 0_{3 \times 3} \\
0_{3 \times 3} & 0_{3 \times 3} & 0_{3 \times 3} & 0_{3 \times 3} \\
0_{3 \times 3} & \mathbf{G}_{3 \times 3} & 0_{3 \times 3} & 0_{3 \times 3}
\end{array}\right]
\end{equation}
where,
\begin{equation}
    G = \begin{bmatrix}
    g sin\psi_0 & g cos\psi_0 & 0\\
    -g cos\psi_0 & g sin\psi_0 & 0\\
    0 & 0 &0
    \end{bmatrix}
\end{equation}
\\
\begin{equation}
B=\left[\begin{array}{cccc}\textbf{0}_\textbf{{6x1}} & \textbf{0}_\textbf{{6x1}} & \textbf{0}_\textbf{{6x1}} & \textbf{0}_\textbf{{6x1}} \\ 0 & \frac{1}{J_{xx}} & 0 & 0 \\ 0 & 0 & \frac{l}{J_{yy}} & 0 \\ 0 & 0 & 0&\frac{1}{J_{zz}} \\ \textbf{0}_\textbf{{2x1}} & \textbf{0}_\textbf{{2x1}} & \textbf{0}_\textbf{{2x1}} & \textbf{0}_\textbf{{2x1}} \\  \frac{1}{m} & 0 & 0 & 0\end{array}\right]
\end{equation}


Controllability matrix is given by
\begin{equation}
    C=\begin{bmatrix}
    B & AB & A^2B . . . . A^{11}B
    \end{bmatrix}
\end{equation}
Substituting mass and inertia values of Heli-quad from Table \ref{parameters}, rank of $C$ comes out to 12. As controllability matrix is full rank, Heli-quad is globally null controllable even under complete failure of an actuator.

\subsection{Outer and Inner loop controllers}

\begin{figure}
\centering
\subfloat[]{\includegraphics [width=\linewidth]{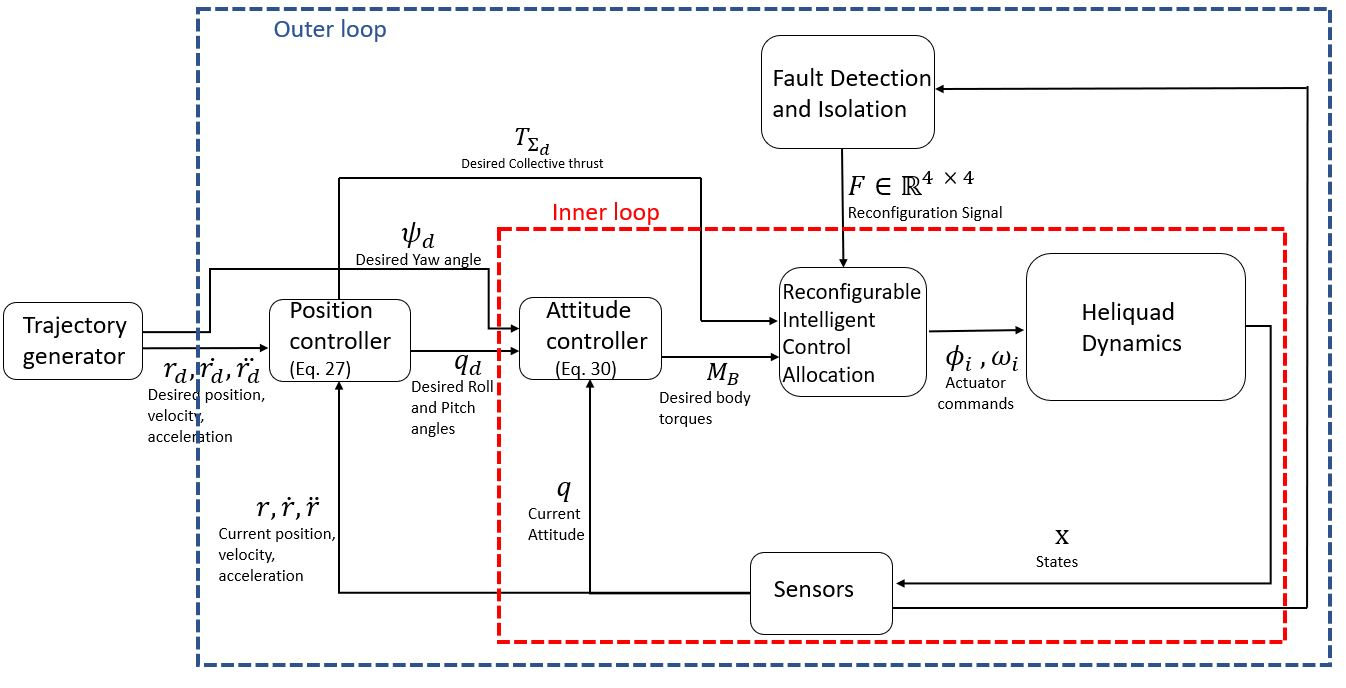}\label{controldiagram}}
 \\
 \subfloat[]{\includegraphics [width=\linewidth]{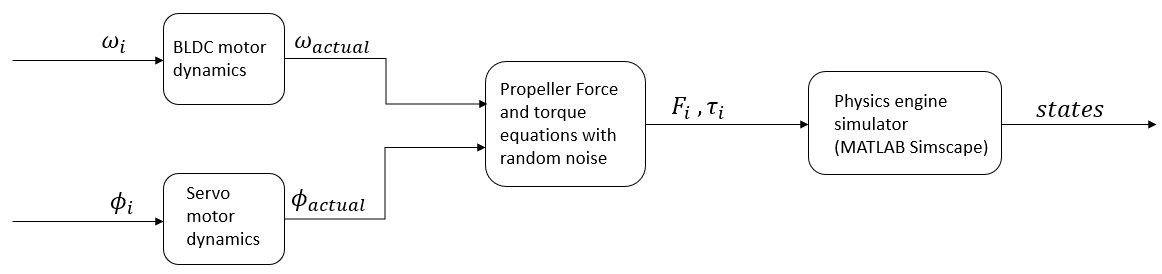}\label{insideHeli-quad}}
 \\
\caption{(a) Overall control diagram for a Heli-quad (b) Modelled Heli-quad dynamics}
\label{schematiccontrol}
\end{figure}

The proposed control architecture is shown in Fig (\ref{controldiagram}) and the block diagram for the modelled dynamics of Heli-quad is shown in Fig (\ref{insideHeli-quad}). The outer loop controller will provide the desired collective thrust vector $\vec{T_{\Sigma_d}}$ based on both the position and velocity errors. In contrast, the inner loop proportional derivative controller will give the desired body torque vector $M_B$ required to track  $\vec{T_{\Sigma_d}}$ and the desired yaw angle. Based on the output of the Fault Detection and Isolation (FDI) routine, which indicates the fault of an  actuator, if any , the intelligent control allocation scheme provides the desired commands for the actuators for a nominal scenario and also for the faulty scenario where the faulty actuator is identified and for the rest of the actuators appropriate torque and thrust commands are generated.

The desired collective thrust vector ($\vec{T_{\Sigma_d}}$) is given by
\begin{eqnarray}
\vec{T_{\Sigma_d}} &=& m \left(p (r_d-r) + d (\dot{r_d}-\dot{r}) + \vec{g} \right) \\
    T_{\Sigma_d}  &=& \| \vec{T_{\Sigma_d}} \| = T_1 + T_2 + T_3 + T_4 
\end{eqnarray}
where, $r=\begin{bmatrix}
 x& y& z
\end{bmatrix}^T$ is current position, $p=diag(\begin{bmatrix}
p_1& p_2& p_3
\end{bmatrix})$ and $d=diag(\begin{bmatrix}
d_1& d_2& d_3 
\end{bmatrix})$ are tuneable gains and $\vec{g}=\begin{bmatrix}
 0 &0 & 9.81
\end{bmatrix}^T$. 

The outputs of inner loop attitude control are the desired body moments given by
  \begin{eqnarray}
  M_B   &=& \begin{bmatrix}
  \tau_x \\ \tau_y \\ \tau_z\\
 \end{bmatrix}_d = \begin{bmatrix}
  (T_4 -T_2 ) l  \\ (T_1 -T_3) l \\ -\tau_1+\tau_2-\tau_3+\tau_4
 \end{bmatrix} \\
 &=& k_p q_e + k_v (\Omega_d-\Omega)    
 \end{eqnarray}
 $q_e$ is the error quaternion vector which represents the angle between $\vec{T_\Sigma}_{d}$ and $Z_B$, $ \Omega_d$ and $\Omega$ are the desired and current body rate vectors respectively, $k_p=diag(\begin{bmatrix}
  k_{p1}&k_{p2}&k_{p3}
 \end{bmatrix})$ and $k_v=diag(\begin{bmatrix}
  k_{v1}&k_{v2}&k_{v3}
 \end{bmatrix})$ are tuneable gains. More details on the inner-loop attitude control can be found in [\citen{cutler}].

\subsection{Reconfigurable fault tolerant control allocation scheme}
In this section, a generalized control allocation scheme is developed to handle both the nominal and faulty conditions. Based on the fault geometry, the neural network based reconfigurable control allocation generates the desired commands for the remaining working actuators. The reconfigurable control allocation approach consists of two steps. In the first step, the desired thrust and torque commands for the individual rotors are calculated. In the second step, there are multiple pitch angles to generate desired thrust, to resolve this actuator redundancy, the desired individual propeller pitch angles ($\phi$) are calculated first then a Neural-Network is implemented to calculate the desired RPM ($\omega$) for a given thrust and torque commands. 

\subsubsection{Calculation of the thrust and torque for individual rotor} \label{calcthrust}

In this paper, it is assumed that there exists a fault detection and isolation scheme, which can identify the failure of an actuator and flag it. Electric current based FDI as given in [\citen{stephan}] can be used to detect the failure of an actuator in a Heli-quad. The index of failed rotor is denoted by $\gamma$. 
\begin{equation}
    \gamma \in \{ 0, 1, 2, 3, 4\}
\end{equation}
$\gamma$ takes the value of zero when there is no fault detected. The FDI will output reconfiguration signal matrix $F$ based on $\gamma$. When there is no fault detected ($\gamma=0$) $F$ will be matrix of ones.
 \begin{equation}
 F=
  \begin{bmatrix} 
\textbf{1}_\textbf{4x4}
\end{bmatrix}   
 \end{equation}
 After the fault is detected, FDI indicates which actuator is faulty ($\gamma$ can be either 1 or 2 or 3 or 4), then based on the value of $\gamma$ some elements of $F$ will be replaced. For example, If $\gamma=3$ then $F$ matrix will assume the value given in Eq.(\ref{gamma=3}).
\begin{equation}
  \begin{bmatrix} 
1&1&0&1\\
1&1&0&1\\
0&1&1/\lambda_3&1\\
0&1&0&1
\end{bmatrix}  
 \label{gamma=3}   
\end{equation}

 The generalized way of replacing elements of $F$ matrix based on $\gamma$ is given below
 \begin{align}
  F_{1\gamma}=F_{2\gamma}=F_{4\gamma}=0\\
      F_{3\gamma}= 1/\lambda_\gamma\\
      F_{3(\gamma\pm2)}=F_{4(\gamma\pm2)}=0 
 \end{align}
 In the above equations $F_{mn}$ denotes the element in $m^{th}$ row and $n^{th}$ column and $\pm$ includes the elements which exists.
 
The output vector $Y \in \mathbb{R}^4$ consisting of desired thrust and torques is defined as
\begin{equation}
 Y=(C \circ F)^{-1} X
 \label{cdotf}
\end{equation}
where $\circ$ denotes element-wise multiplication of matrices (Hadamard product) , matrix $C$ represents the control allocation matrix in nominal operation and the vector $X\in\mathbb{R}^4$ is the output of attitude controller.
\begin{equation}
 C=
\begin{bmatrix} 
0&-l&0&l\\
l&0&-l&0\\
-\lambda_1&\lambda_2&-\lambda_3&\lambda_4\\
1&1&1&1
\end{bmatrix}
\end{equation}
\begin{equation}
X =  [M_B \quad T_{\Sigma_D}]^T 
 \end{equation}

 During nominal operation,  the output vector $Y$ will consist of individual thrust forces. During faulty operation,  $Y$ will consist of desired thrust forces of three working motors and desired torque of rotor opposite to failed one ($Y_\gamma=\tau_{\gamma\pm2}$).
 For example, when $\gamma=3$
\begin{align}
    Y &= 
    \left(\begin{bmatrix}
    0&-l&0&l\\
l&0&-l&0\\
-\lambda_1&\lambda_2&-\lambda_3&\lambda_4\\
1&1&1&1
    \end{bmatrix} \circ    \begin{bmatrix} 
1&1&0&1\\
1&1&0&1\\
0&1&1/\lambda_3&1\\
0&1&0&1
\end{bmatrix}\right)^{-1}X \\
\\
&= \left(\begin{bmatrix}
0&-l&0&l\\
l&0&0&0\\
0&\lambda_2&-1&\lambda_4\\
0&1&0&1
\end{bmatrix}\right)^{-1}X\\
\\
Y&=\begin{bmatrix}
T_1 \quad  T_2 \quad \tau_1 \quad T_4
\end{bmatrix}^T
\end{align}

Note that the matrix ($C \circ F$) in Eq.(\ref{cdotf}) is full rank for all the fault scenarios.

\subsubsection{Calculation of actuator commands} \label{calcactuator}
Multiple pitch angles can achieve desired thrust command because of actuator redundancy in the Heli-quad. In the nominal flight conditions, propeller pitch can be varied according to the desired flight mode. The hover pitch angle should be set at the lower side for aggressive flights while keeping the RPM constant. It is beneficial to set the propeller pitch on the higher side for power-efficient operation throughout the flight. In this paper, for nominal flight conditions, propeller pitch is set to a  constant value at the higher side ($\phi_{nom}=4 deg$) on all the four actuators meaning $\lambda$ is also the same for all the actuators. Setting the propeller at this pitch angle gives enough control authority for nominal flight (see Fig.\ref{thrusttorque}).

\begin{equation}
 \therefore  \lambda_1=\lambda_2=\lambda_3=\lambda_4=\lambda_{nom}  \quad \textrm{when $\gamma=0$} \\
\end{equation}

For faulty conditions, only two opposite working actuators should generate the thrust for lifting the vehicle. For simplicity, the propeller pitch angles on these two actuators are fixed to a constant value. The torque produced by these actuators should be minimum as their combined torque will be counteracted by a single actuator, opposite to the failed one. However, as seen in Fig.\ref{pitchlambda}, $\lambda$ doesn't vary significantly at higher pitch angles, hence setting pitch angles at higher values generates a least torque for producing the desired thrust. In this paper, propeller pitch on two opposite working actuators are set at $\lambda_{fault} = 10deg$.
\begin{equation}
  \therefore \lambda_{\gamma\pm1}=\lambda_{\gamma\pm3}=\lambda_{fault} \  \textrm{when $\gamma\neq0$}     
\end{equation}
Desired thrust and torque values for an actuator opposite to the failed one will be given by output vector $Y$ as discussed in  section \ref{calcthrust}. The desired pitch angle for this actuator is calculated as shown below.
\begin{equation}
    \lambda_{\gamma\pm2}=Y_\gamma /Y_{\gamma\pm2}
\end{equation}
\begin{equation}
    \phi_{\gamma\pm2}=f_\lambda^{-1}(\lambda_{\gamma\pm2})
\end{equation}
In simulations, a look-up-table is used to approximate function $f_{\lambda}^{-1}$. Once the desired pitch angles for all actuators are known, a fully connected neural network is used to get desired RPM commands. 

\begin{figure}
\centering
\includegraphics [width=0.9\linewidth]{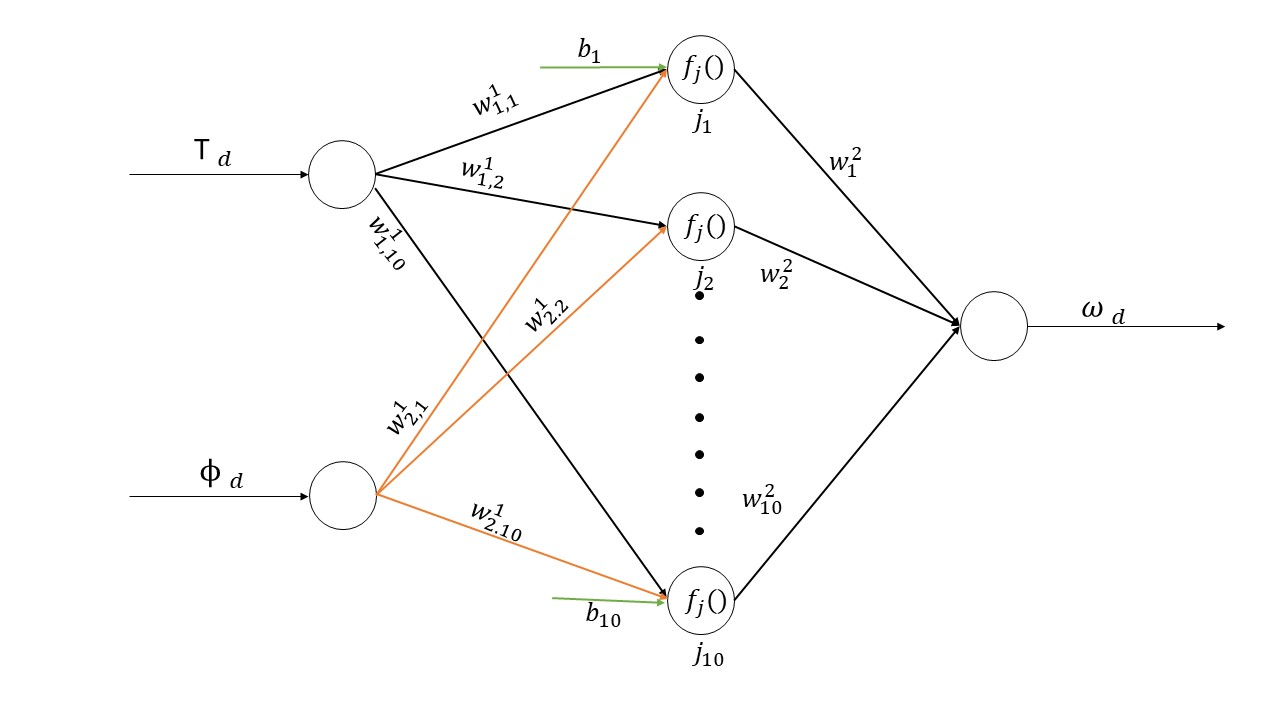}
\caption{Neural network architecture for control allocation. There are 10 neurons in hidden layer.}
\label{nnc}
\end{figure}

Thrust ($T$) and torque ($\tau$) are  a function of $\phi$ and $\omega$ and are given by
\begin{eqnarray}
    T(\phi,\omega) &=& k_1(\phi) \omega^2 + k_2(\phi) \omega + k_3(\phi) \\
    \tau(\phi,\omega) &=& q_1(\phi) \omega^2 + q_2(\phi) \omega + q_3(\phi)
\end{eqnarray}
where $k_1,k_2,k_3,q_1,q_2,q_3$ are some nonlinear functions in $\phi$ [\citen{cutler}]. In order to handle the nonlinearities in propeller aerodynamics, a control allocation neural network is proposed which approximates the relationship between $x=[T_d,~\phi_d]^T$ and the desired speed ($\omega_d$) for the actuator. First time in the literature, a neural network is used in control allocation for VPP systems. However, the use of Neural-Network to handle the complex aerodynamic effects in the fixed-pitch quadcopter control allocation has been experimentally validated in [\citen{nncite}]. The architecture of the control allocation neural network used in this paper is shown in Fig. \ref{nnc} with $N_H$ number of hidden neuron. The approximated $\omega_d$ is given as
\begin{equation}
    \hat{\omega}_d=\sum^{N_H}_{k=1} w^{2}_{k} h_k
\end{equation}
where 
\begin{equation}
    h_k=f \left(\sum^2_{i=1} w^{1}_{ik} x_i+ b_{k}\right), ~k=1,2,\cdots,N_H
\end{equation}
$f(.)$ is a tanhyperbolic activation function. \linebreak

\begin{figure}
\centering
\includegraphics [width=\linewidth]{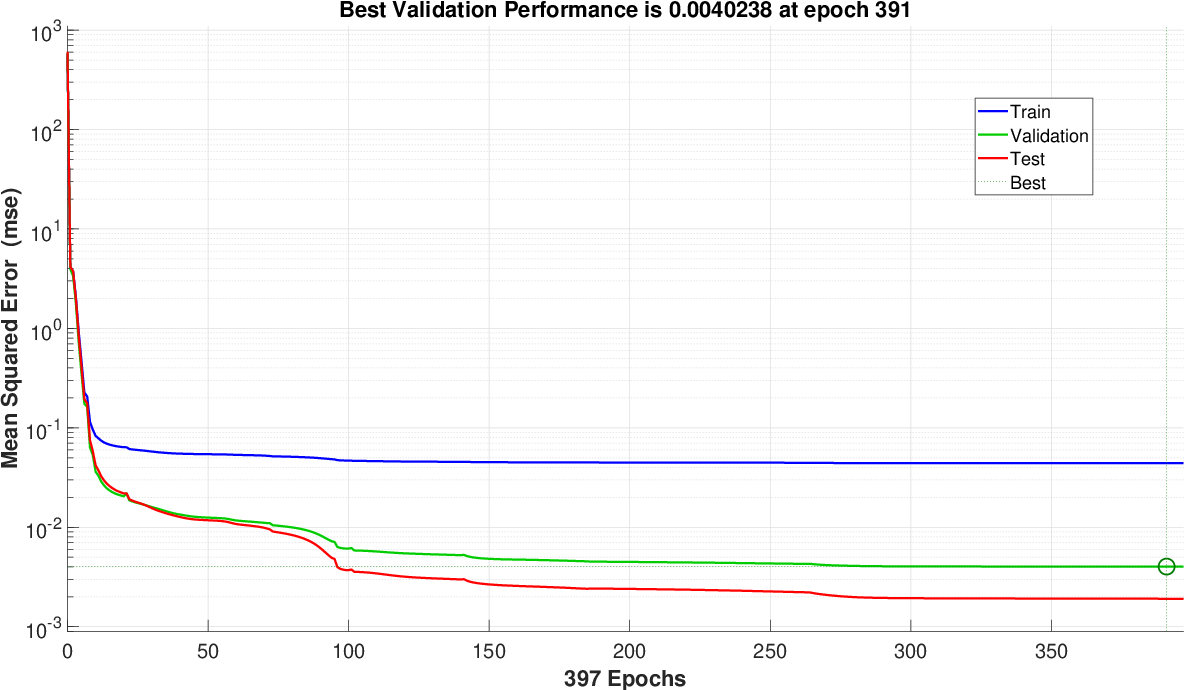}
\caption{Neural network performance.}
\label{nnperf}
\end{figure}

The training data for the control allocation network is generated using Blade Element Momentum Theory for different number of inputs $\phi$ and $\omega$. Standard levenberg-marquardt backpropogation algorithm was used to train the weights ($w$) and biases ($b$). The number of hidden neurons were selected using the procedure described in [\citen{vijay}]. Training was conducted for 1000 epochs. Fig. \ref{nnperf} shows the performance of the neural network. It can be seen that the network is able to approximate the data very well.

\section {Performance Evaluation of the Fault-Tolerant Heli-quad Controller} \label{performanceeval}
In this section, the Heli-quad’s full attitude fault-tolerant capability is evaluated. High fidelity software in the loop simulations were conducted to analyze the Heli-quad’s performance in both nominal as well as faulty conditions. The simulations are conducted using MATLAB Simulink and Simscape environments running on the Intel i7-8600 processor with 8GB RAM. First, the simulation model development is described followed by the controller’s tracking performance for the nominal flight condition. Next, the Heli-quad’s tracking performance under a complete failure of a single actuator is highlighted. Finally, the effect of the response time for the FDI to detect the fault on the system’s stability is also discussed. 

\subsection {Simulation model development}
\begin{figure}
\centering
\includegraphics [width=0.9\linewidth]{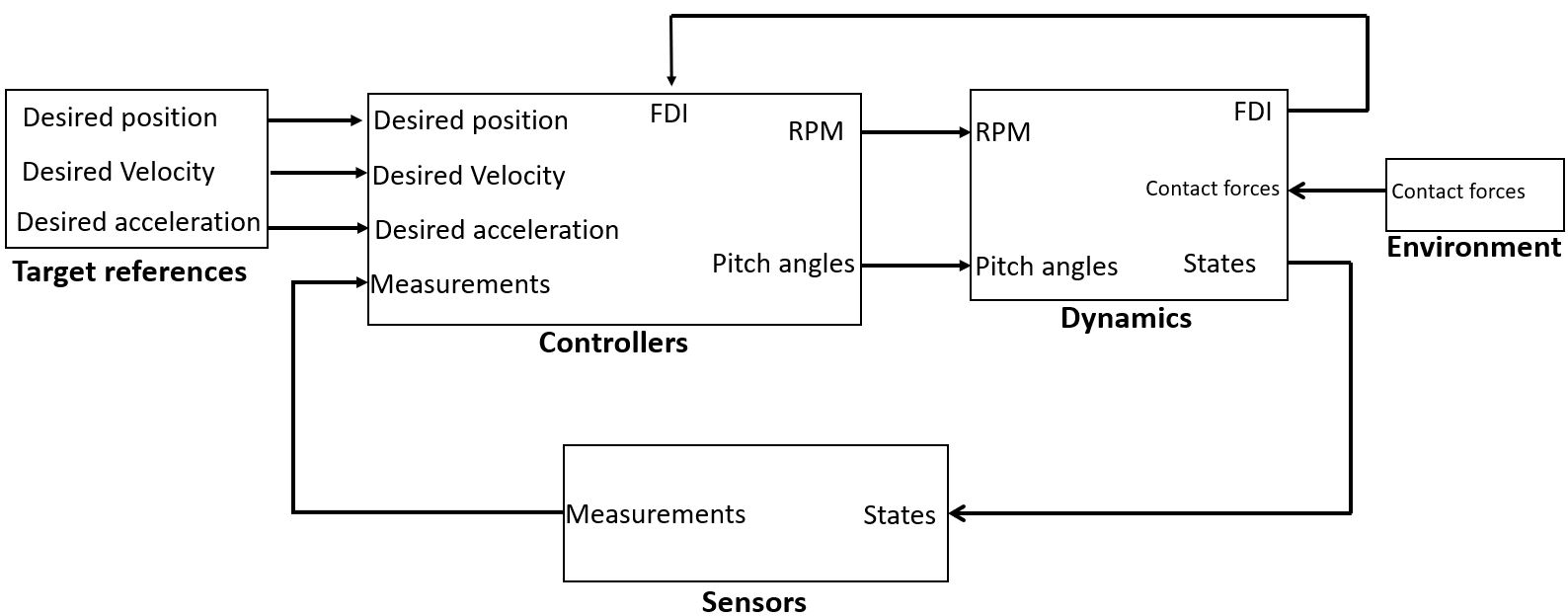}
\caption{Overview of the Simulink model used in the simulations.}
\label{simulink}
\end{figure}

In this subsection, the developed simulation model which includes actuator dynamics, noise, disturbance modelling, and Heli-quad’s dynamics are explained in detail. The overall simulation code written in Simulink environment is shown in Fig.\ref{simulink}. Inside the 'Target References'  block, the waypoints are interpolated by a time-parameterized cubic polynomial to output the desired position, velocities, and acceleration. The 'Controllers' block contains the position and attitude controller along with the control allocation algorithm as discussed in the section \ref{intelftcontroldesign}. The 'Dynamics' block contains the actuator dynamics and the Heli-quad’s 6-DOF dynamics which is modelled in the Simscape environment. This block will also output the FDI signal that will be fed to the 'Controllers' block for reconfiguring the control allocation loop after a failure. The 'Sensors' block placed in the feedback path models the sensor noise based on the actual state values. The 'Environment' block contains the Simscape code for modelling the floor (ground) and contact forces between it and the Heli-quad.

In reality, the actual output from an actuator may not be the same as the desired one. Heli-quad has two actuators and one to control RPM of the propeller and  a digital servo motor control the blade pitch angle. The BLDC motor is modelled by standard equations available in the literature [\citen{cutler}].
\begin{equation}
 I\dot{\omega} = \left(\left(V-\frac{\omega}{K_v}\right)\frac{1}{R} - I_0 \right)\frac{1}{K_v} - \tau
 \label{motoreq}
\end{equation}
where $V$ is the input voltage, $I_0$ is the no-load current of the motor, $R$ is the internal resistance, and $K_v$ is the voltage constant for the motor. $\tau$ is the opposing torque acting on the motor. These parameter values used for the simulation are given in Table \ref{parameters}.

 \begin{table}
\centering
\caption{Heli-quad inertial and BLDC motor parameters}\label{parameters}
{\begin{tabular}{|c|c|c|}
\hline
Parameters &Value& Units \\ 
\hline
$l$ & 0.1794 &  m\\  
\hline
 $J_{XX}$ & $3.34 \times 10^{-2}$ & $Kg-m^2$\\
 \hline
$J_{YY}$ & $3.34 \times 10^{-2}$ & $Kg-m^2$\\
\hline
$J_{ZZ}$ & $6.66 \times 10^{-2}$ & $Kg-m^2$\\
\hline
$V$ & 12 & Volts\\ 
\hline
$K_v$ & 1000 & rad/s-V\\
\hline
$R$  & 0.09 & Ohms\\
\hline
$I_0$ & 0.5 & Amp\\ 
\hline
\end{tabular}}
\end{table} 

The servo motor dynamics is much faster than the BLDC motor dynamics and can be modelled as just a rate limit of $500~deg/sec$.

In real life scenarios, for a given pitch angle and RPM of the propeller, the actual thrust and torque values generated by it may not match exactly with the values predicted by BEMT. The differences arise due to many causes such as backlash in the mechanism, modelling uncertainty in the aerodynamic parameters, external wind, etc. As discussed in section \ref{experimental}, BEMT was able to predict the thrust and torque values within $\pm 10$ percent of the actual values. To make the simulations more realistic, $10$ percent of random noise is added to the theoretical values as given in Eq.\ref{tnoise} and Eq.\ref{taunoise} . 
 \begin{equation}
\begin{split}
    & T(\phi_i,\omega_i)= T(\phi_i,\omega_i)_{BEMT} + w,\quad \text{where} \\
    & w \sim U[-0.1 \ T(\phi_i,\omega_i)_{BEMT},\ 0.1 \ T(\phi_i,\omega_i)_{BEMT}]
    \end{split}
    \label{tnoise}
\end{equation}
\begin{equation}
\begin{split}
& \tau(\phi_i,\omega_i)= \tau(\phi_i,\omega_i)_{BEMT} + u,\quad \text{where} \\
& u  \sim U[ \ -0.1 \ \tau(\phi_i,\omega_i)_{BEMT} ,\ 0.1 \ \tau(\phi_i,\omega_i)_{BEMT}]
    \end{split}
    \label{taunoise}
\end{equation}
 
\begin{figure}
\centering
\subfloat[]{\includegraphics [width=0.9\linewidth]{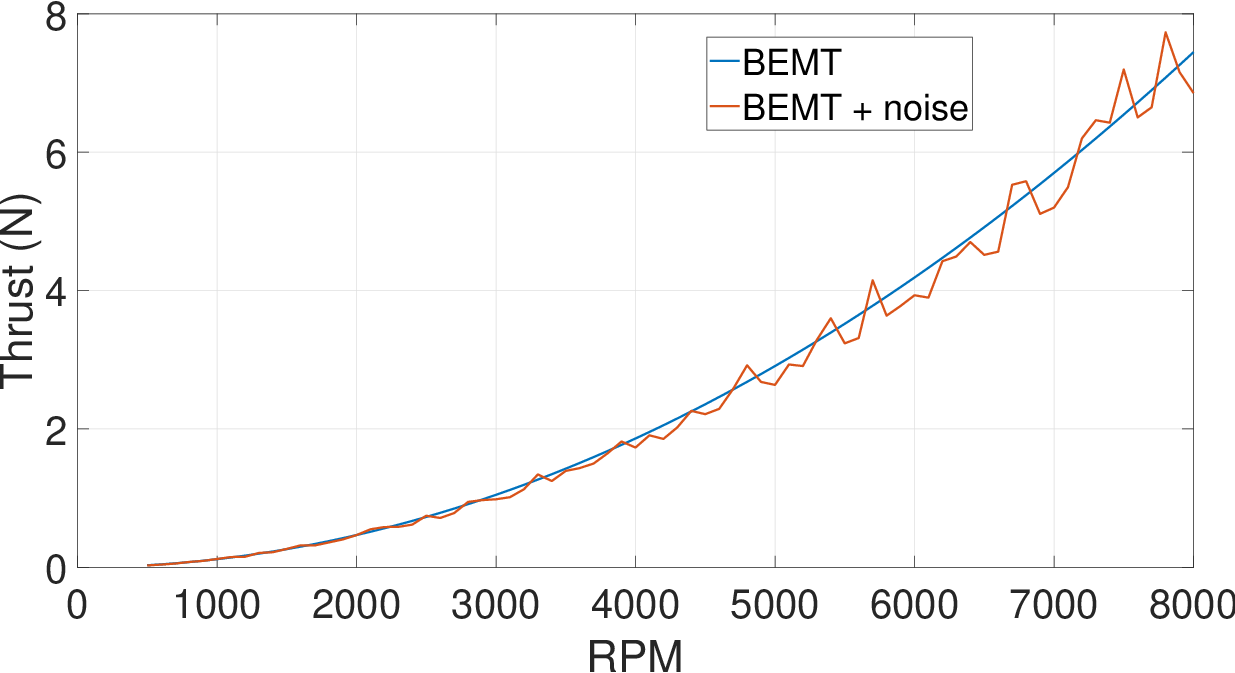}}
\\
 \subfloat[]{\includegraphics [width=0.9\linewidth]{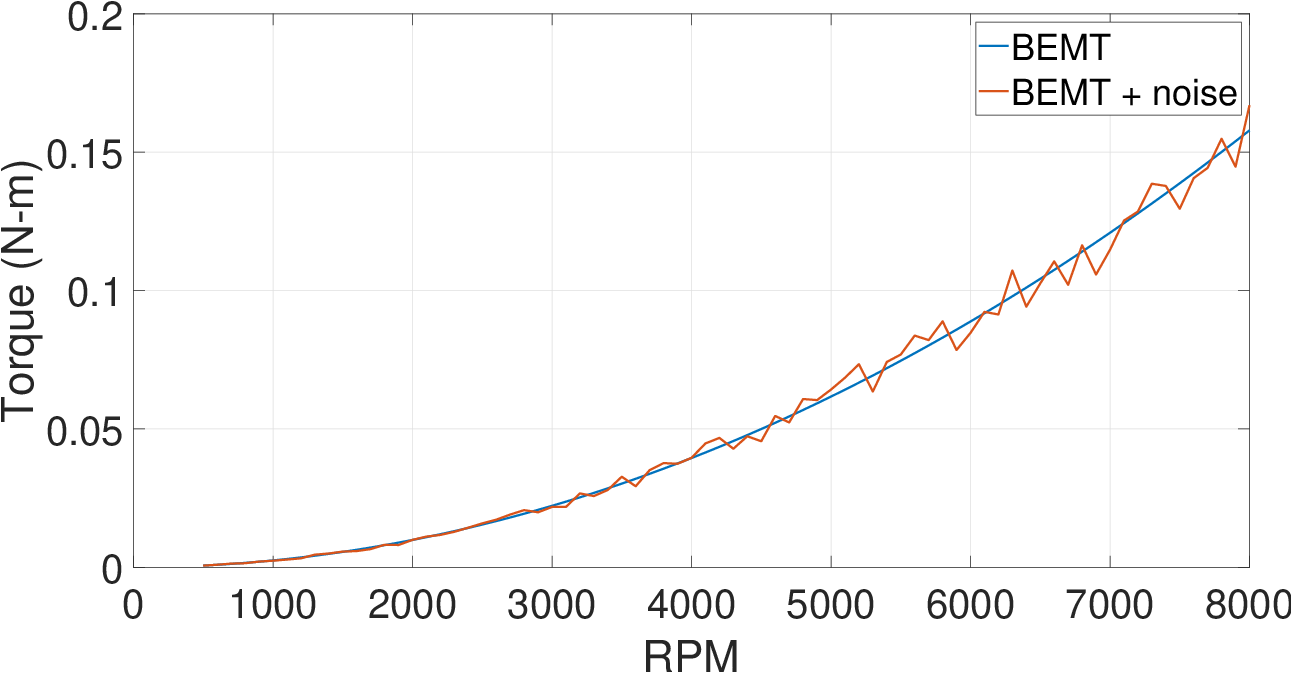}}
\caption{(a) Propeller thrust force with noise at 4 degree pitch angle (b) Propeller torque with noise at -2.6 ($\phi_{T=0}$) degree pitch angle}
\label{noise}
\end{figure}

In the above equations, $T/\tau(\phi_i,\omega_i)_{BEMT}$ are the values predicted by BEMT. $U[a,b]$ denotes the random number from uniform distribution between the interval $a$ and $b$. $u$ and $w$ assumes different values everytime the simulation is executed. Fig.\ref{noise} shows the difference between the BEMT and realistic thrust/torque values used in the simulations for one pitch angle and for one instance. Overall, the difference between the BEMT and simulated thrust/torque values are within $\pm 10 \%$ for all instances.
 
\begin{figure}
\centering
\includegraphics [width=\linewidth]{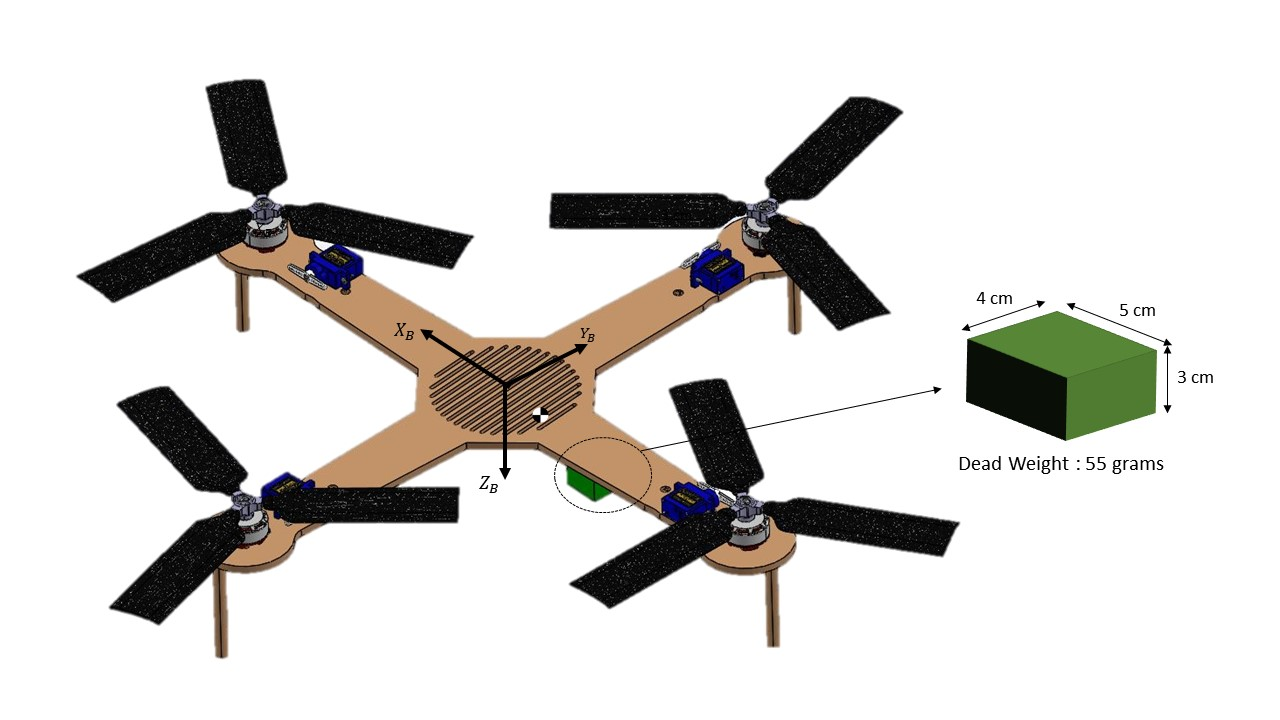}
\caption{CAD model of Heli-quad used in the simulations. Additional dead weight is added to shift Centre of Gravity from the centroid of the geometry.}
\label{vehicle}
\end{figure}

The 6DOF dynamics of Heli-quad is simulated in MATLAB Simscape multi-body environment \footnote{\url{https://in.mathworks.com/products/simscape-multibody.html}}. Simscape multibody formulates and solves the equations of motion with very high accuracy based on a Unified Robot Description Format (URDF) model of the robot. Fig.\ref{vehicle} shows the CAD model used to generate the URDF code for the Heli-quad.

Some important physical parameters of the URDF model are given in Table \ref{parameters}. In real life scenarios, the Centre Of Mass (COM) will not be exactly at the centroid of the geometry. To simulate this offset in COM an additional deadweight is added. The vector from centroid to COM described in the body axes is given by $d_{B,COG}=[-2\quad -0.7 \quad  0.5]^T cms$.
To simulate the sensor, white Gaussian noise with zero mean and stand deviation of $20$ cms for all the three position coordinates and $0.5$ deg for each angle were added to the actual values [\citen{sensornoise}].

\subsection{Heli-quad’s performance for a nominal flight condition}
\begin{figure}
\centering
\includegraphics [width=0.9\linewidth]{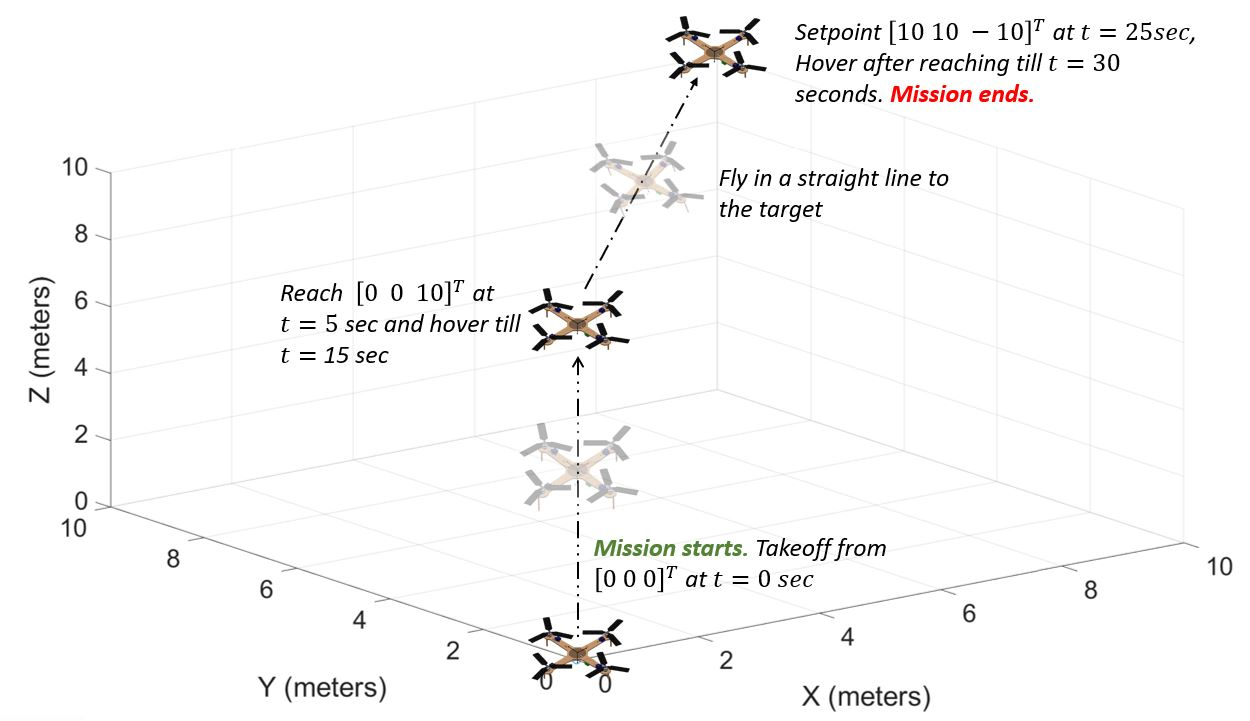}
\caption{Waypoints at various time intervals used to demonstrate simulations. Sign of $Z$ on the graph is ignored for clarity.}
\label{waypoint}
\end{figure}

In this subsection, Heli-quad’s nominal performance under uncertainties and sensor noise is evaluated. There is no fault induced in any of the Heli-quad’s actuators throughout the simulations. Fig.\ref{waypoint} shows the waypoints given for the simulation. The simulation starts when the Heli-quad is resting at the origin. At $t=0$, it is commanded to climb to an altitude of 10 meters in 5 seconds and stay there, maintaining the same inertial $X$ and $Y$ coordinates. At $t=15$ seconds, while the Heli-quad is hovering, the desired inertial $X$ and $Y$ coordinates are changed to $10$ meters each keeping a constant altitude. These coordinates are maintained till the end of the simulation. Yaw angle is commanded to be zero throughout the mission trajectory.  A cubic time parameterized polynomial has been fitted between two successive waypoints to get the desired position, velocity, and acceleration at each timestep.

\begin{figure}
\centering
\includegraphics [width=0.9\linewidth]{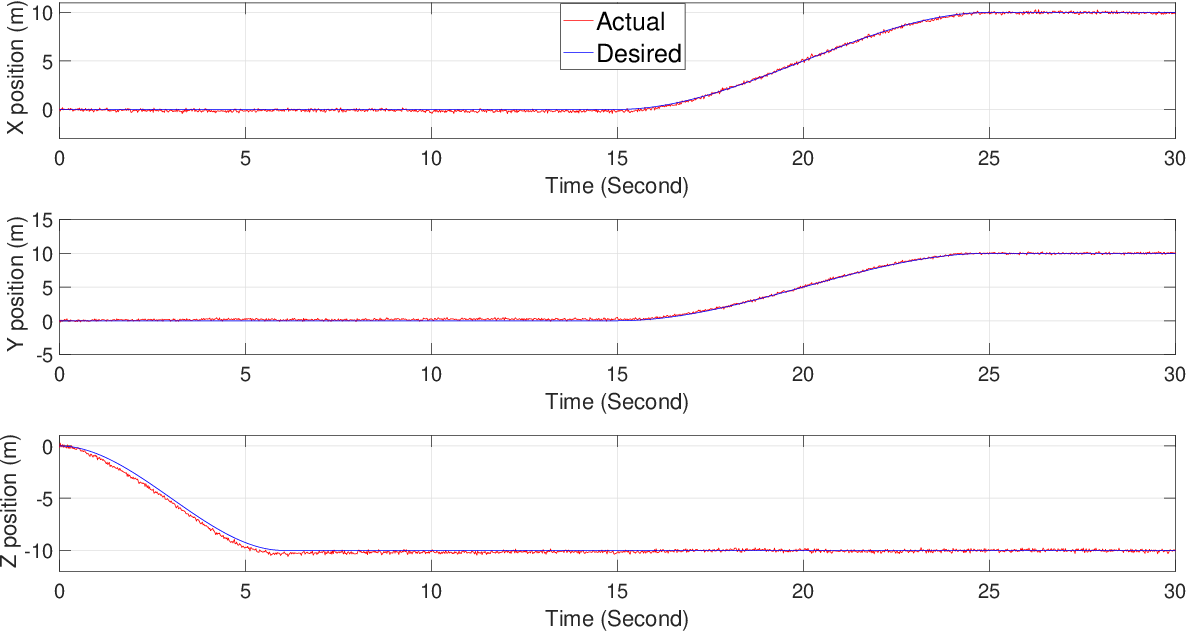}
\caption{Position tracking performance of the Heli-quad (without any failure)}
\label{tvsxyz_nom}
\end{figure}

\begin{figure}
\centering
\includegraphics [width=0.9\linewidth]{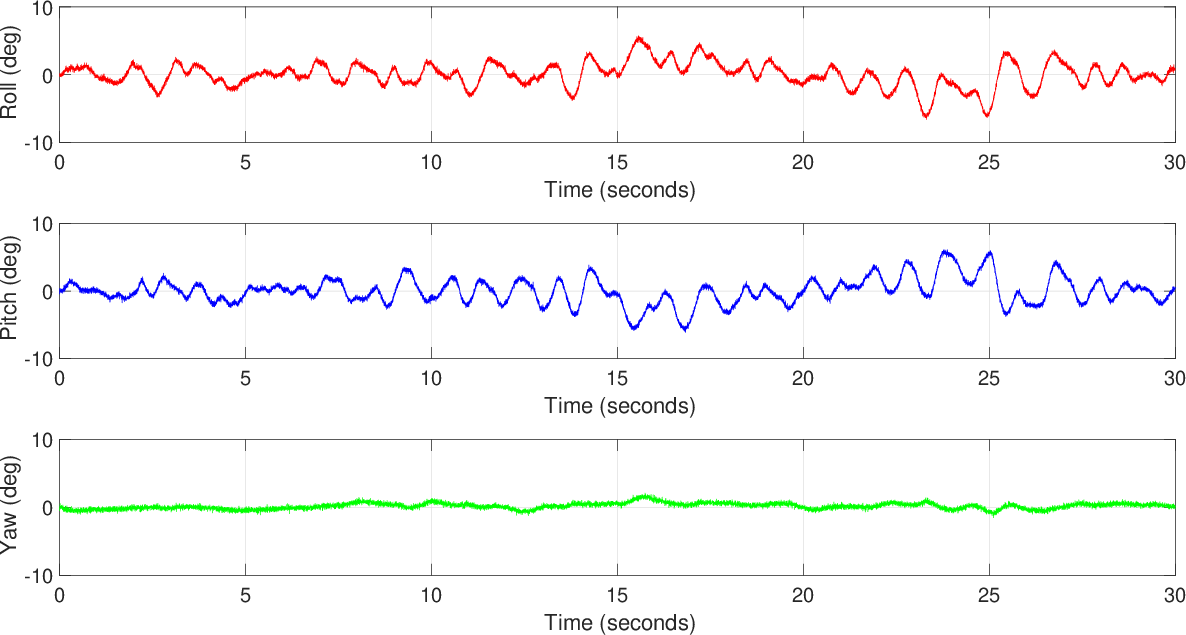}
\caption{Attitude variation during the mission }
\label{tvsattitude_nom}
\end{figure}

Fig.\ref{tvsxyz_nom} shows the position tracking capability of the Heli-quad in a fault-free flight. It can be seen that the Heli-quad is able to track the desired position with good accuracy even under uncertainties and sensor noise. The actual position coordinates followed by the Heli-quad are always within approximately $0.5$ meters of the desired path. Fig.\ref{tvsattitude_nom} shows the attitude variation of the Heli-quad. All the angles are within acceptable ranges.\linebreak

\begin{figure}
\centering
\includegraphics [width=\linewidth]{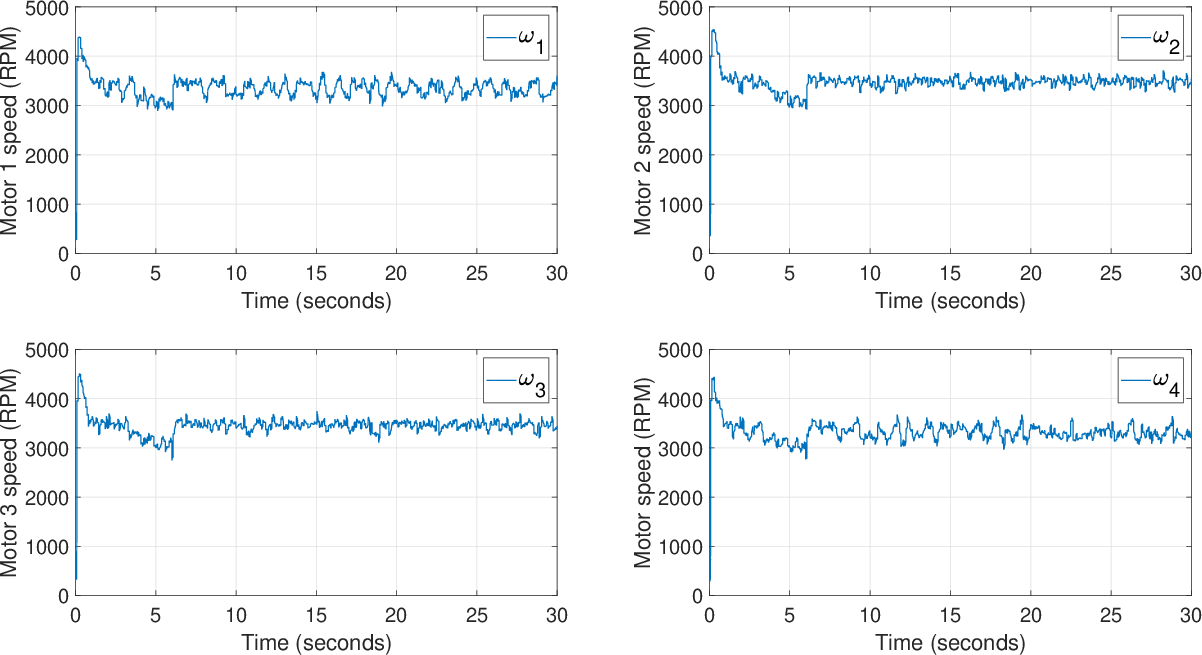}
\caption{RPM commands variation for all the motors}
\label{tvsRPM_nom}
\end{figure}

The variation of commanded RPM is shown in Fig.\ref{tvsRPM_nom}. Note, as mentioned in the section \ref{calcactuator}, the pitch angle is kept constant in the nominal fault-free flight and in these simulations, it is kept constant at $4$ degrees. It can be seen that the commanded RPMs of all the motors are within the acceptable bounds. The results show that the controller for the Heli-quad can track the position references with adequate accuracies even under dynamic uncertainties and sensor noise.

\subsection{Heli-quad’s performance under the complete failure of a single actuator}
The performance of the Heli-quad when one of it’s actuators gets completely compromised (one motor fails and does not rotate) is presented. The effect of the time delay that FDI takes to detect the fault on Heli-quad’s stability is also discussed. To simulate the instantaneous complete failure of an actuator, zero RPM command is given to the respective BLDC motor. 

The desired waypoints for this simulation are the same as described in the nominal case. However, at $t=12$ seconds, when the Heli-quad is hovering, the BLDC motor of actuator $4$ ($\gamma$=4) is turned off to simulate the sudden failure \footnote{The simulation results for failure occurring during the forward flight are given in the supplementary material}. It is assumed that FDI takes $50$ milliseconds to detect the fault. As a result of this delay, the control allocation matrix won’t be reconfigured till $t=12.05$ seconds. To evaluate the full attitude controllability, after the failure, the desired yaw angle is set to the constant value of $-25$ degrees for the rest of the trajectory. 
 
 \begin{figure}
\centering
\subfloat[12 Seconds]{\includegraphics[width=0.3\linewidth]{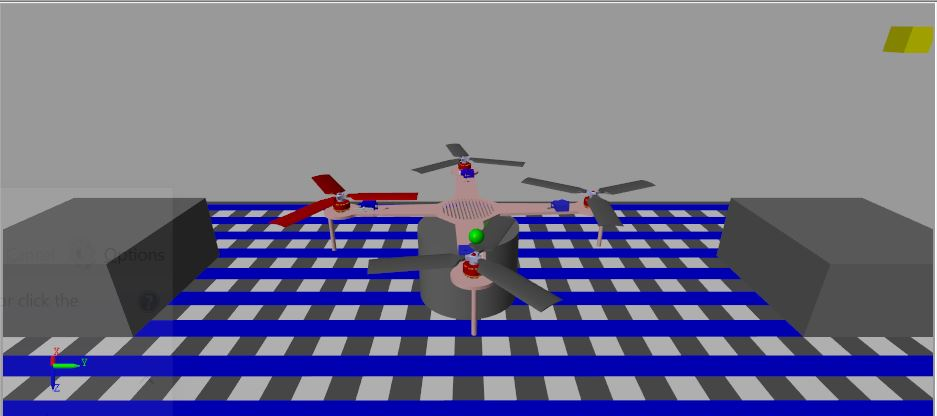}}
\subfloat[12.2 Seconds]{\includegraphics[width=0.3\linewidth]{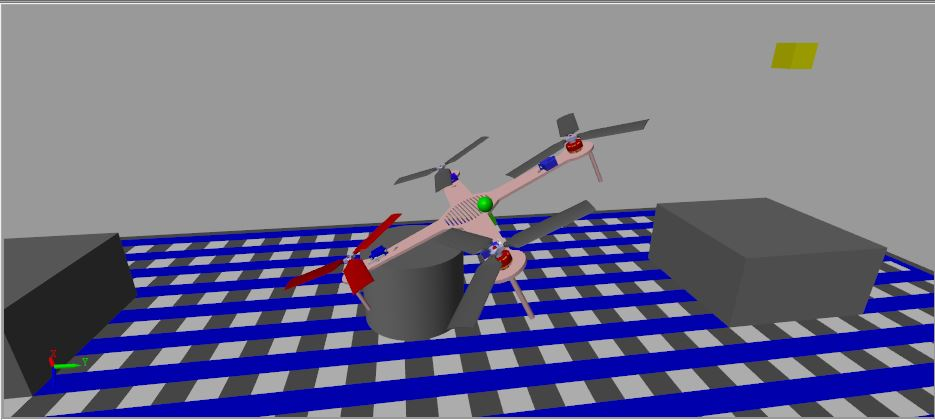}}
\subfloat[12.3 Seconds]{ \includegraphics[width=0.3\linewidth]{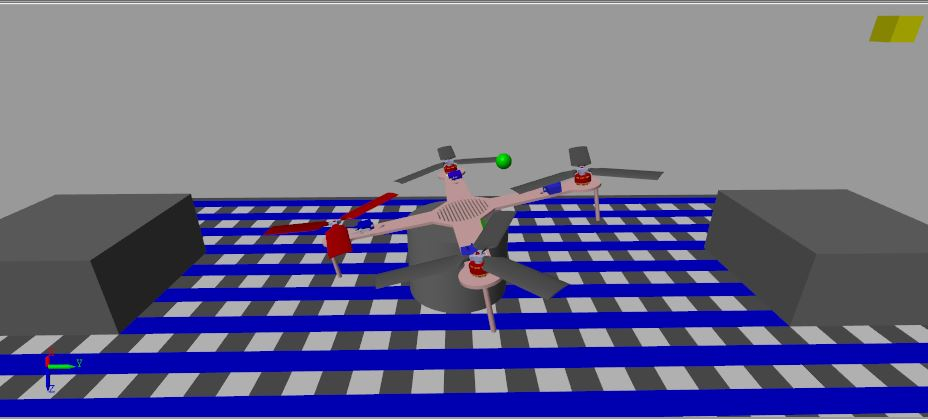}}
\\
\subfloat[12.5 Seconds]{\includegraphics[width=0.3\linewidth]{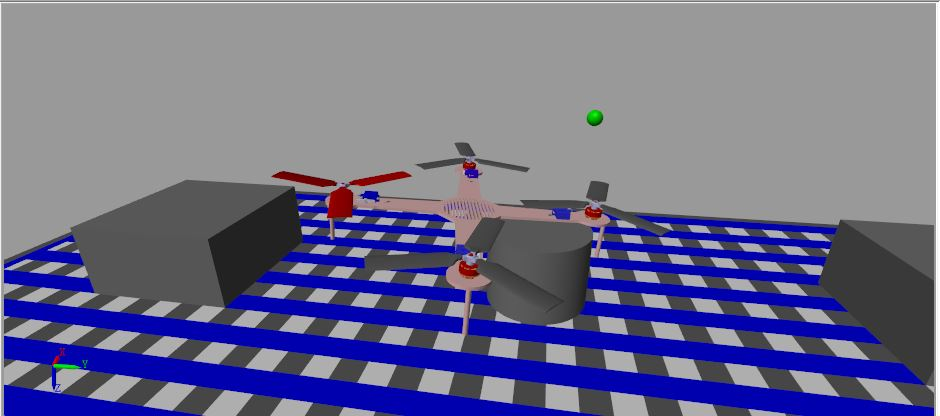}}
\subfloat[12.7 Seconds]{\includegraphics[width=0.3\linewidth]{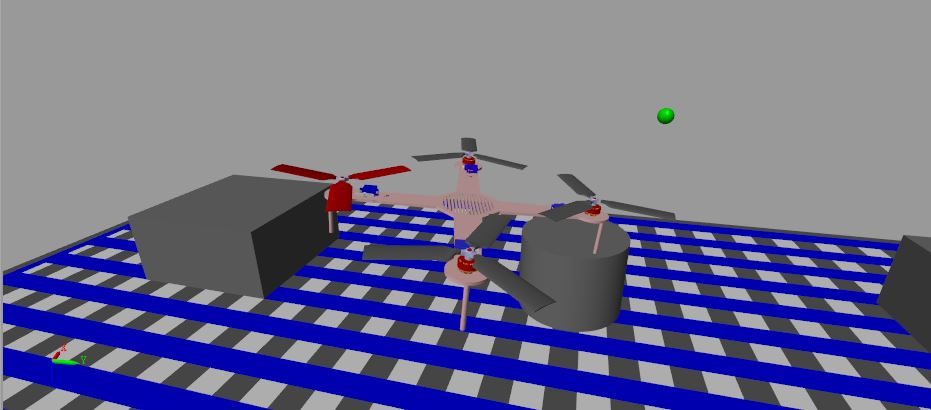}}
\subfloat[13 Seconds]{\includegraphics[width=0.3\linewidth]{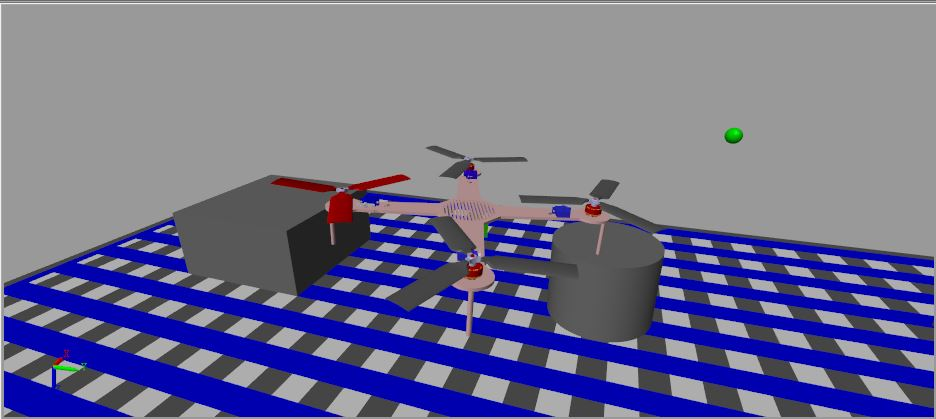}}
 \caption{Simulation snapshots at different time interval. Red colour propeller stops rotating at 12sec, fault is detected at $12.05$sec. Heli-quad levels back at around 12.7 seconds. }
\label{simscape}
\end{figure}

\begin{figure}
\centering
\includegraphics [width=0.9\linewidth]{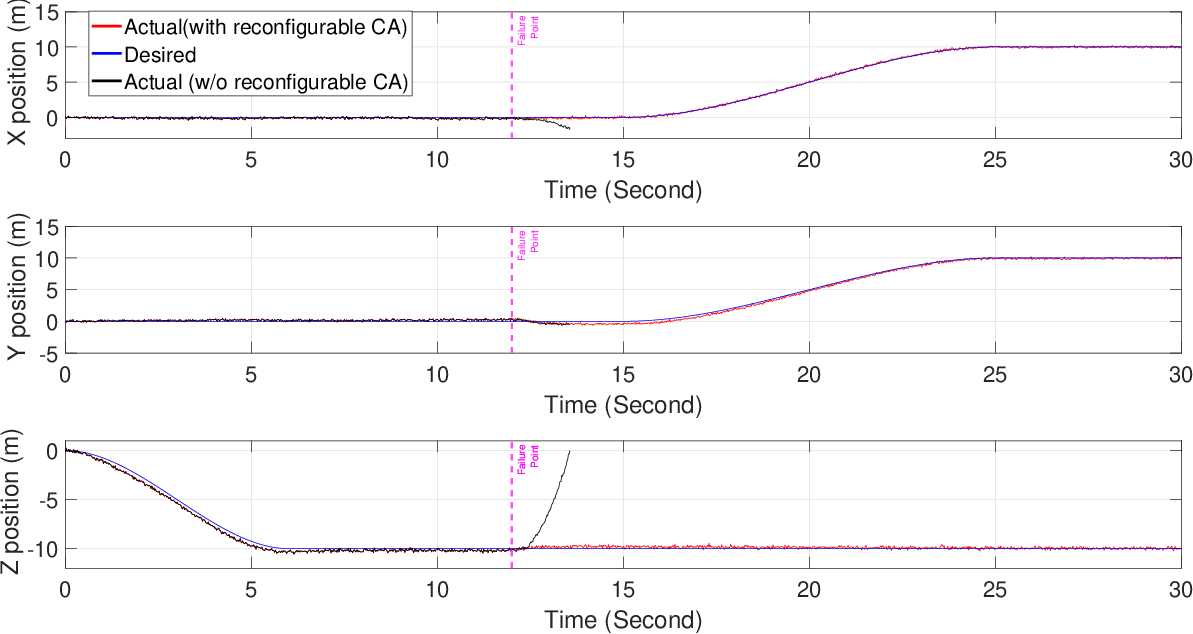}
\caption{Position tracking performance of the Heli-quad under complete failure of an actuator. Dashed magenta line at $t=12$ indicates the time of failure.}
\label{tvsxyz}
\end{figure}

The snapshots of the Simscape simulations just after the failure are given in Fig.\ref{simscape} \footnote{The simulation video is given with the supplementary material}. As soon as the red colour propeller stops spinning, the Heli-quad tilts due to the thrust imbalance. However it is able to recover back to near-hover state after the failure, in approximately $0.7$ seconds.  Fig.\ref{tvsxyz} shows the position tracking capability of the Heli-quad under the complete failure of actuator 4. After failure, when using the reconfigurable control allocation, it can be seen that due to FDI delay there is some increase in the tracking error. However, the error is within 1 meter of the desired value. After the desired waypoints are changed at t = 15 seconds; it can be seen that the Heli-quad is able to track the position references with sufficient accuracies using only the remaining three fully working actuators. In sharp contrast, the Heli-quad exponentially loses altitude after failure when a non-reconfigurable control allocation (similar to a fixed-pitch quadcopter) is used.\linebreak

\begin{figure*}
\centering
\includegraphics [width=0.7\linewidth]{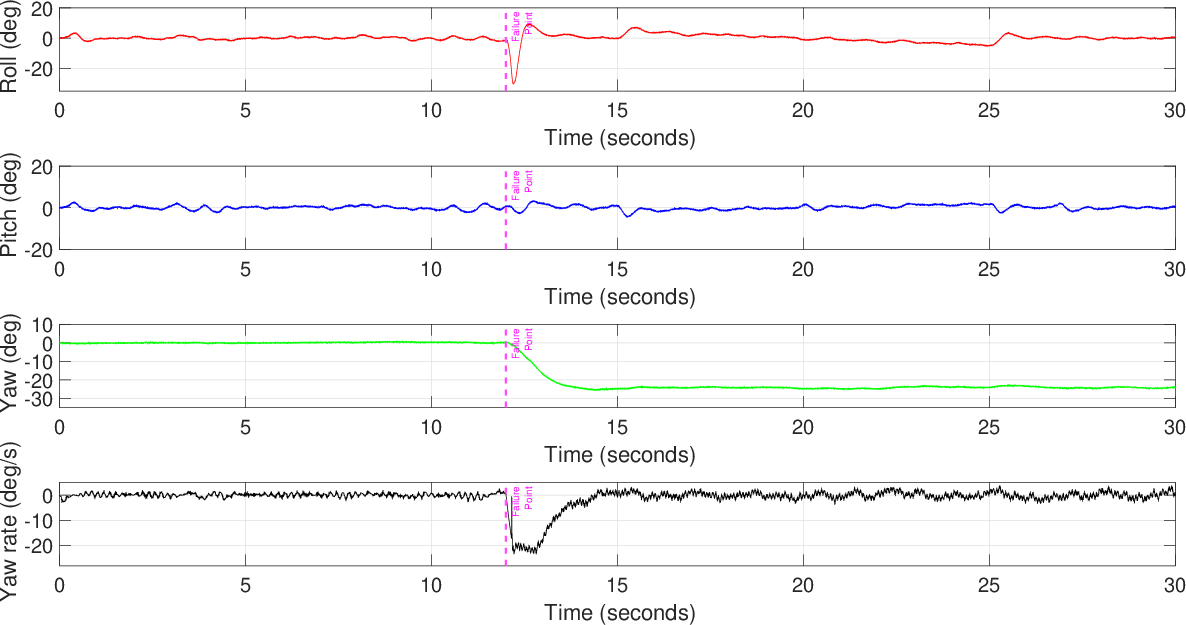}
\caption{Attitude variation during the mission. Note, after failure yaw-rate goes to zero.}
\label{tvsattitude}
\end{figure*}

Fig.\ref{tvsattitude} shows the Heli-quad’s attitude and yaw rate variations. After the failure of the actuator $4$ at $t=12$ seconds, due to the FDI delay, the thrust imbalance causes Heli-quad’s roll angle magnitude to increases to $27$ degrees before the control allocation is reconfigured. However, it can be seen that the controller is able to bring the Heli-quad back to zero roll angle. Due to the FDI time delay, torque imbalance also makes the Heli-quad to spin in the clockwise direction as soon as the failure occurs, but, after the reconfiguration of control allocation, the unique ability of Heli-quad to bring the yaw rate back to zero can be clearly seen. The controller is also able to track the desired yaw angle with acceptable accuracy.

 \begin{figure*}
\centering
\includegraphics [width=0.8\linewidth]{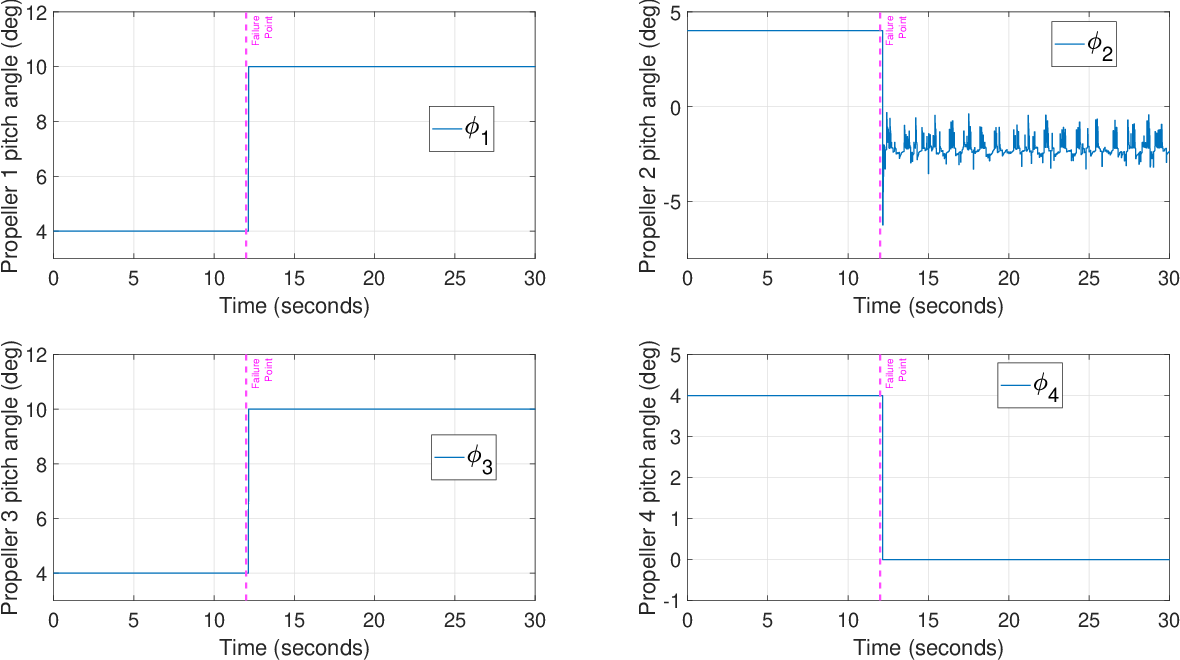}
\caption{Propeller pitch angle variation during mission.}
\label{tvspitch}
\end{figure*}

 \begin{figure*}
\centering
\includegraphics [width=0.8\linewidth]{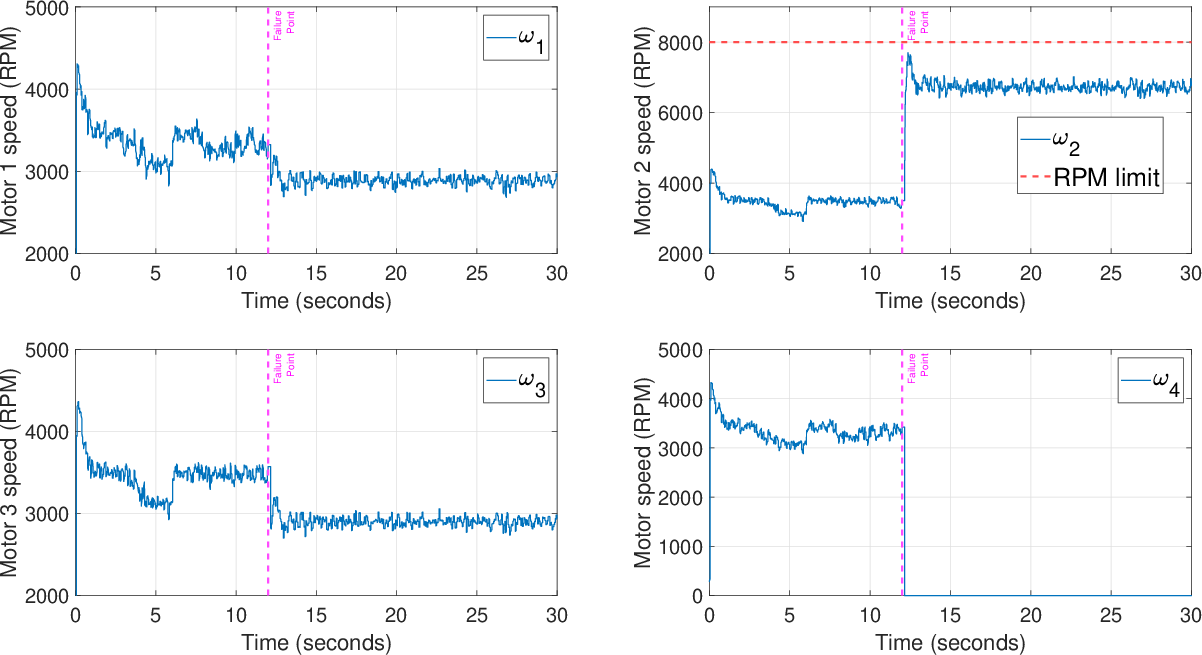}
\caption{RPM of all the motors. Motor 4 is turned off to simulate failure.}
\label{tvsrpm}
\end{figure*}

Fig.\ref{tvspitch} and Fig.\ref{tvsrpm} shows the commands given to the actuators. Before the failure, the pitch angles of all four actuators are set constant at $4$ degrees. In this phase, only the RPM of each motor is varied to track references. As soon as the fault is detected, as discussed in the section \ref{calcactuator}, the pitch angle of the actuators $1$ and $3$ are set to a constant value of $10$ degrees. Note, the controller commands the pitch angle of propeller $2$ ($\phi_2$) to assume a negative value to counteract the adverse roll angle by generating negative thrust. After stabilizing back to hover, $\phi_2$ remains near $\phi_{T=0}$ of $-2.6$ degrees throughout the flight. The RPM of motors $1$ and $3$ decreases slightly after the failure, this happens because of the increase in the pitch angle on those actuator. In contrast, after failure, in order to nullify the adverse yaw rate, the RPM of the actuator $2$ increases from its nominal value of $3000$ to $6500$. In these RPM ranges the torque generated by it is equal to/greater than the torque generated by propellers $1$ and $3$ combined. It can be seen the RPMs of all the motors are within reasonable limits.

From all the figures given above, it can be clearly inferred that the time it takes for the FDI to detect the fault is crucial for the Heli-quad’s stability. An increase in this fault detection time delay increases adverse attitude rates making it harder for the Heli-quad to recover. To estimate the maximum possible time delay that could be handled by the controller, multiple simulations with varying time delays were conducted. It has been found that a maximum time delay from which the Heli-quad is able to successfully recover is $180$ milliseconds. 
 
\section{Conclusion} \label{conclusion}

This paper has presented a new variable pitch quadcopter with a cambered airfoil propeller called Heli-quad that can control all the attitude angles even under a single actuator complete failure. Experimental test data shows that it is not feasible to control the yaw angle using nominal symmetric airfoil propellers because of the inadequate torque produced at those pitch angles that generate a zero thrust.  The use of a cambered airfoil propeller design provides the necessary yaw control authority by generating a torque of $4.5$ times that of its symmetric airfoil counterpart at zero -a thrust pitch angle. The experimental test data also shows the existence of a hover equilibrium point for the Heli-quads weighing a maximum of up to $815$ grams. Controllability analysis of the linear dynamics of the Heli-quad guarantees its hover stability with just three working actuators. The unified control architecture that includes the novel reconfigurable control allocation scheme is shown to make Heli-quad’s controller tractable under arbitrary actuator failure. High-fidelity software-in-the-loop simulations that contain the sensor noise and various uncertainties indicate the performance of the Heli-quad in nominal fault-free flight by achieving a $0.31$ meter RMS error in position tracking. Two different simulations with the faults occurring during hover and forward flight are also conducted in the faulty flight cases. In both cases, the results clearly indicate that the proposed Heli-quad can handle the single actuator failure and provide necessary position tracking performances with an  RMS error of $0.37$ and $0.41$ meters respectively while maintaining a constant yaw angle. In all the simulations, it is assumed that it takes 50 milliseconds by the FDI scheme to detect the fault. With this time delay, it takes around $0.7$ seconds for the Heli-quad to return to the level attitude. The effect of fault detection times on the Heli-quad’s stability have been found to be crucial. Detailed studies carried out with different FDI delays indicate that the maximum possible time delay that the Heli-quad’s controller can handle is $180$ milliseconds (with $50$ grams payload).

\bibliographystyle{ieeetr}
\bibliography{Bibliography}

\begin{thebibliography}{10}

\bibitem{hari20}
K.~Harikumar, J.~Senthilnath, and S.~Suresh, ``Mission aware motion planning (map) framework with physical and geographical constraints for a swarm of mobile stations,'' {\em IEEE Transactions on Cybernetics}, vol.~50, p.~1209–1219, 03 2020.

\bibitem{hari18}
K.~Harikumar, J.~Senthilnath, and S.~Suresh, ``Multi-uav oxyrrhis marina-inspired search and dynamic formation control for forest firefighting,'' {\em IEEE Transactions on Automation Science and Engineering}, vol.~16, pp.~863--873, 02 2018.

\bibitem{agriculture}
P.~{Katsigiannis}, L.~{Misopolinos}, V.~{Liakopoulos}, T.~K. {Alexandridis}, and G.~{Zalidis}, ``An autonomous multi-sensor uav system for reduced-input precision agriculture applications,'' in {\em 2016 24th Mediterranean Conference on Control and Automation (MED)}, (Athens,Greece), pp.~60--64, 2016.

\bibitem{ben_search}
J.~Q. Cui, S.~K. Phang, K.~Z. Ang, F.~Wang, X.~Dong, Y.~Ke, S.~Lai, K.~Li, X.~Li, J.~Lin, {\em et~al.}, ``Search and rescue using multiple drones in post-disaster situation,'' {\em Unmanned Systems}, vol.~4, no.~01, pp.~83--96, 2016.

\bibitem{geology}
B.~Jordan, ``A bird’s-eye view of geology: The use of micro drones/uavs in geologic fieldwork and education,'' {\em GSA Today}, vol.~25, pp.~42--43, 07 2015.

\bibitem{construction}
S.~Siebert and J.~Teizer, ``Mobile 3d mapping for surveying earthwork projects using an unmanned aerial vehicle (uav) system,'' {\em Automation in construction}, vol.~41, pp.~1--14, 2014.

\bibitem{senthil2021}
J.~Senthilnath, A.~Kumar, A.~Jain, K.~Harikumar, M.~Thapa, S.~Suresh, G.~Anand, and J.~A. Benediktsson, ``Bs-mcl: Bilevel segmentation framework with metacognitive learning for detection of the power lines in uav imagery,'' {\em IEEE Transactions on Geoscience and Remote Sensing}, vol.~60, pp.~1--12, 2021.

\bibitem{landry}
B.~Landry, R.~Deits, P.~R. Florence, and R.~Tedrake, ``Aggressive quadrotor flight through cluttered environments using mixed integer programming,'' in {\em 2016 IEEE international conference on robotics and automation (ICRA)}, pp.~1469--1475, IEEE, 2016.

\bibitem{mohta}
K.~Mohta, M.~Watterson, Y.~Mulgaonkar, S.~Liu, C.~Qu, A.~Makineni, K.~Saulnier, K.~Sun, A.~Z. Zhu, J.~Delmerico, K.~Karydis, N.~Atanasov, G.~Loianno, D.~Scaramuzza, K.~Daniilidis, C.~J. Taylor, and V.~Kumar, ``Fast, autonomous flight in gps-denied and cluttered environments,'' {\em J. Field Robotics}, vol.~35, pp.~101--120, 2018.

\bibitem{airtaxi}
C.~Silva, W.~R. Johnson, E.~Solis, M.~D. Patterson, and K.~R. Antcliff, ``Vtol urban air mobility concept vehicles for technology development,'' in {\em 2018 Aviation Technology, Integration, and Operations Conference}, (Atlanta, Georgia, USA), p.~3847, 2018.

\bibitem{pavel}
M.~D. Pavel, ``Understanding the control characteristics of electric vertical take-off and landing (evtol) aircraft for urban air mobility,'' {\em Aerospace Science and Technology}, vol.~125, p.~107143, 2021.

\bibitem{partiallossofblades}
M.~I. Alabsi and T.~D. Fields, ``Real-time closed-loop system identification of a quadcopter,'' {\em Journal of Aircraft}, vol.~56, no.~1, pp.~324--335, 2019.

\bibitem{partialfailure}
B.~Wang and Y.~Zhang, ``Adaptive sliding mode fault-tolerant control for an unmanned aerial vehicle,'' {\em Unmanned Systems}, vol.~5, no.~04, pp.~209--221, 2017.

\bibitem{Nguyenpartialfailure}
N.~P. Nguyen and S.~K. Hong, ``Fault-tolerant control of quadcopter uavs using robust adaptive sliding mode approach,'' {\em Energies}, vol.~12, p.~95, Dec 2018.

\bibitem{taes}
J.~I. Giribet, R.~S. Sanchez-Pena, and A.~S. Ghersin, ``Analysis and design of a tilted rotor hexacopter for fault tolerance,'' {\em IEEE Transactions on Aerospace and Electronic Systems}, vol.~52, no.~4, pp.~1555--1567, 2016.

\bibitem{hexacopter}
M.~E. McKay, R.~Niemiec, and F.~Gandhi, ``Analysis of classical and alternate hexacopter configurations with single rotor failure,'' {\em Journal of Aircraft}, vol.~55, no.~6, pp.~2372--2379, 2018.

\bibitem{parachute}
B.~Al-Madani, M.~Svirskis, G.~Narvydas, R.~Maskeli{\=u}nas, and R.~Dama{\v{s}}evi{\v{c}}ius, ``Design of fully automatic drone parachute system with temperature compensation mechanism for civilian and military applications,'' {\em Journal of Advanced Transportation}, vol.~2018, pp.~1--11, 2018.

\bibitem{freddi}
A.~Freddi, A.~Lanzon, and S.~Longhi, ``A feedback linearization approach to fault tolerance in quadrotor vehicles,'' {\em IFAC Proceedings Volumes}, vol.~44, no.~1, pp.~5413 -- 5418, 2011.
\newblock 18th IFAC World Congress.

\bibitem{mueller}
M.~W. Mueller and R.~D'Andrea, ``Stability and control of a quadrocopter despite the complete loss of one, two, or three propellers,'' in {\em 2014 IEEE international conference on robotics and automation (ICRA)}, pp.~45--52, IEEE, 2014.

\bibitem{cutler}
M.~Cutler and J.~P. How, ``{Analysis and Control of a Variable-Pitch Quadrotor for Agile Flight},'' {\em Journal of Dynamic Systems, Measurement, and Control}, vol.~137, no.~10, 2015.
\newblock 101002.

\bibitem{baldini}
A.~Baldini, R.~Felicetti, A.~Freddi, S.~Longhi, and A.~Monteri{\`u}, ``Actuator fault tolerant control of variable pitch quadrotor vehicles,'' {\em IFAC-PapersOnLine}, vol.~53, no.~2, pp.~4095--4102, 2020.

\bibitem{wang}
Z.~Wang, R.~Groß, and S.~Zhao, ``Control of centrally-powered variable pitch propeller quadcopters subject to propeller faults,'' {\em Aerospace Science and Technology}, vol.~120, p.~107245, 2022.

\bibitem{wang2020}
Z.~Wang, R.~Gro{\ss}, and S.~Zhao, ``Controllability analysis and controller design for variable-pitch propeller quadcopters with one propeller failure,'' {\em Advanced Control for Applications: Engineering and Industrial Systems}, vol.~2, no.~2, p.~e29, 2020.

\bibitem{chipade}
V.~S. Chipade, Abhishek, M.~Kothari, and R.~R. Chaudhari, ``Systematic design methodology for development and flight testing of a variable pitch quadrotor biplane vtol uav for payload delivery,'' {\em Mechatronics}, vol.~55, pp.~94--114, 2018.

\bibitem{chipade2018advanced}
V.~S. chipade, A.~Abhishek, and M.~Kothari, ``Advanced flight dynamic modelling of variable pitch quadrotor,'' in {\em 2018 AIAA Atmospheric Flight Mechanics Conference}, (Kissimmee, Florida, USA), pp.~1763--1779, 2018.

\bibitem{shastry}
A.~K. Shastry, M.~Kothari, and A.~Abhishek, ``Generalized flight dynamic model of quadrotor using hybrid blade element momentum theory,'' {\em Journal of Aircraft}, vol.~55, no.~5, pp.~2162--2168, 2018.

\bibitem{leishman}
J.~G. Leishman, {\em Principles of helicopter aerodynamics}.
\newblock Cambridge University Press, 2006.

\bibitem{kim}
K.~Kim, S.~Rahili, X.~Shi, S.-J. Chung, and M.~Gharib, ``Controllability and design of unmanned multirotor aircraft robust to rotor failure,'' in {\em AIAA Scitech 2019 Forum}, (San Diego, California, USA), p.~1787, 2019.

\bibitem{nullcntrl}
E.~B. Lee and L.~Markus, ``Foundations of optimal control theory,'' tech. rep., Minnesota Univ Minneapolis Center For Control Sciences, 1967.

\bibitem{stephan}
V.~Stepanyan, K.~S. Krishnakumar, and A.~Bencomo, ``Identification and reconfigurable control of impaired multi-rotor drones,'' in {\em AIAA Guidance, Navigation, and Control Conference}, (San Diego, California, USA), p.~1384, 2016.

\bibitem{nncite}
S.~P. Madruga, A.~H. Tavares, S.~O. Luiz, T.~P. do~Nascimento, and A.~M.~N. Lima, ``Aerodynamic effects compensation on multi-rotor uavs based on a neural network control allocation approach,'' {\em IEEE/CAA Journal of Automatica Sinica}, vol.~9, no.~2, pp.~295--312, 2021.

\bibitem{vijay}
M.~Kumar, S.~Omkar, R.~Ganguli, P.~Sampath, and S.~Suresh, ``Identification of helicopter dynamics using recurrent neural networks and flight data,'' {\em Journal of the American Helicopter Society}, vol.~51, no.~2, pp.~164--174, 2006.

\bibitem{sensornoise}
A.~Razinkova and H.-C. Cho, ``Tracking a moving ground object using quadcopter uav in a presence of noise,'' in {\em 2015 IEEE International Conference on Advanced Intelligent Mechatronics (AIM)}, (Busan, South Korea), pp.~1546--1551, 2015.

\end{thebibliography}

\noindent\includegraphics[width=1in]{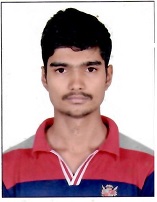}
{\bf Eeshan Kulkarni}  is a Ph.D. student at the Department of Aerospace Engineering, Indian Institute of Science, Bengaluru. His research interests include Unmanned Aerial Vehicle design, control systems design, and field robotics. \\

\noindent\includegraphics[width=1in]{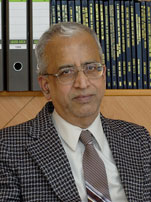}
{\bf Narasimhan Sundararajan} received the Ph.D. degree in electrical engineering from the University of Illinois at Urbana–Champaign, Urbana, IL, USA, in 1971.

From 1971 to 1991, he was with the Vikram Sarabhai Space Centre, Indian Space Research Organization, Trivandrum, India. Since 1991, he has been a Professor (Retd.) with the School of Electrical and Electronic Engineering, Nanyang Technological University (NTU), Singapore. Currently, he is a technical consultant in WIRIN project at IISc Bengaluru, India. His current research interests include spiking neural networks, neuro-fuzzy systems, and optimization with swarm intelligence. \\

\noindent\includegraphics[width=1in]{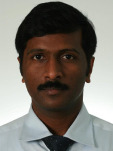}
{\bf Suresh Sundaram} received the Ph.D. degree in aerospace engineering from the Indian Institute of Science, Bengaluru, India, in 2005.

He is currently an Associate Professor with the department of Aerospace Engineering, Indian Institute of Science. From 2010 to 2018, he was an Associate Professor with the School of Computer Science and Engineering, Nanyang Technological University, Singapore. His research interests include flight control, unmanned aerial vehicle design, machine learning, optimization, and computer vision.\\

\end{document}